\documentclass[twocolumn,pre,showpacs,eqsecnum]{revtex4}

\usepackage{dcolumn}
\usepackage{amsfonts}
\usepackage{amsmath}
\usepackage{amssymb}
\usepackage{bm}

\newif\ifpdf
\ifx\pdfoutput\undefined
\pdffalse 
\else
\pdfoutput=1 
\pdftrue
\fi

\ifpdf
\usepackage[pdftex]{graphicx}
\else
\usepackage{graphicx}
\fi

\begin{document}

\ifpdf
\DeclareGraphicsExtensions{.jpg,.pdf,.tif}
\else
\DeclareGraphicsExtensions{.eps,.jpg}
\fi

\newcommand{\brm}[1]{\bm{{\rm #1}}}

\title{Percolating granular superconductors}

\author{Hans-Karl Janssen}
\affiliation{
Institut f\"{u}r Theoretische Physik 
III\\Heinrich-Heine-Universit\"{a}t\\Universit\"{a}tsstra{\ss}e 1\\
40225 D\"{u}sseldorf\\
Germany
}

\author{Olaf Stenull}
\affiliation{
Department of Physics and Astronomy\\
University of Pennsylvania\\
Philadelphia, PA 19104\\
USA
}
\vspace{10mm}
\date{\today}

\begin{abstract}
\noindent
We investigate diamagnetic fluctuations in percolating granular superconductors. Granular superconductors are known to have a rich phase diagram including normal, superconducting and spin glass phases. Focusing on the normal-superconducting and the normal-spin glass transition at low temperatures, we study  he diamagnetic susceptibility $\chi^{(1)}$ and the mean square fluctuations of the total magnetic moment $\chi^{(2)}$ of large clusters. Our work is based on a random Josephson network model that we analyze with the powerful methods of renormalized field theory. We investigate the structural properties of the Feynman diagrams contributing to the renormalization of $\chi^{(1)}$ and $\chi^{(2)}$. This allows us to determine the critical behavior of $\chi^{(1)}$ and $\chi^{(2)}$  to arbitrary order in perturbation theory. 
\end{abstract}

\pacs{64.60.Ak, 74.80.Bj, 74.25.Fy}

\maketitle

\section{Introduction} 
\noindent
The discovery of high-temperature superconductivity raised hope of an important role of this phenomenon in applications requiring high current densities, such as power transmission lines and high-field magnets. The most severe factor limiting current densities is that all practical materials contain defects like impurities, grain boundaries and other extended defects. Thus, it is important to investigate the role of disorder in superconductivity. 

Percolation theory~\cite{bunde_havlin_91_etc} plays a predominant role in the study of disordered systems. In the course of the years it has been applied to granular superconductors in many ways~\cite{zeimets_etal_02}. The diamagnetic properties of disordered composites of superconducting and non-superconducting materials can by studied in terms of a percolation model where superconducting grains are located on the sites of a ($d$-dimensional) hypercubic lattice and where Josephson junctions occupy nearest neighbor bonds  with a given probability $p$. In the following we will refer to such a network as a random Josephson network (RJN).

The phase diagram of the RJN has a rich structure depending on the occupation probability $p$, the temperature $T$, and an external magnetic field $B$. This phase diagram was explored in a seminal work by John and Lubensky (JL)~\cite{john_lubensky_85_86}. Viewing the unoccupied bonds as normal conductors, one has a normal phase,  a Meissner phase, and a spin-glass (SG) phase. For type II superconducting materials one has in addition an Abrikosov flux lattice phase.  For $p$ below the percolation threshold $p_c$ there are only finite superconducting clusters and hence there is no macroscopic superconductivity. For $p$ exceeding $p_c$ there exists at least one spanning cluster and the phase depends on $B$. As $p \to p_c$ at $B=0$ one encounters for sufficiently low temperatures a transition from the normal to the Meissner phase (henceforth, we refer to this transition as transition I). This Meissner phase has its typical hallmarks, viz., expulsion of magnetic flux and  a non-vanishing average condensate wave function. For $B>0$ the system crosses over between the normal and the SG phase as $p_c$ is approached at low $T$ (this transition will be referred to from now on as transition II). The SG phase is characterized by a vanishing average condensate wave function but non-vanishing Edwards-Anderson order parameter~\cite{edward_anderson_75}. 

Early theoretical work on the transition I dates back to the beginning 1980's. It was predicted by de Gennes~\cite{deGennes_81} and Alexander~\cite{alexander_83_etc} that the configurationally averaged diamagnetic susceptibility $\chi^{(1)}$ diverges as
\begin{subequations} 
\label{prediction}
\begin{eqnarray}
\label{prediction_a}
\chi^{(1)} \sim \left| p - p_c  \right|^{-\varphi}
\end{eqnarray}
with
\begin{eqnarray}
\label{prediction_b}
\varphi = 2 \nu - t \, ,
\end{eqnarray}
\end{subequations}
where $\nu$ is the percolation correlation length exponent and $t$ ist the conductivity exponent of the random resistor network (RRN). A few years later John, Lubensky, and Wang (JLW)~\cite{JLW_88} presented a renormalization group analysis of the RJN based on a replicated Landau-Ginzburg-Wilson (LGW) type continuum model introduced in Ref.~\cite{john_lubensky_85_86}. They studied $\chi^{(1)}$ as well as the mean square fluctuations $\chi^{(2)}$ of the total magnetic moment for transitions I and II. Their results obtained in a perturbation calculation to one-loop order support the prediction~(\ref{prediction}). Moreover, their calculations resulted in $\chi^{(2)} = 0$ for transition I and $\chi^{(1)} = 0$ for transition II. $\chi^{(2)}$ was found to diverge at transition II as
 \begin{eqnarray}
\label{schrott}
\chi^{(2)} \sim \left| p - p_c  \right|^{t_2 - 2\nu} \, ,
\end{eqnarray}
where $t_2$ is a crossover exponent distinct from $t$. Roux and Hansen~\cite{roux_hansen_88} carried out numerical work to calculate $\varphi$ for $d=2$ dimensions. As JLW, Roux and Hansen relied on linearized network equations. Their result $\varphi = 1.36 \pm 0.02$ agrees well with the scaling relation (\ref{prediction_b}) if the established values for $\nu$ and $t$ are inserted. Wang and Lubensky~\cite{wang_lubensky_91} utilized a low concentration series expansion to determine $\varphi$ for $d=2$. Their result $\varphi = 1.21 \pm 0.03$, however, is inconsistent with the scaling relation (\ref{prediction_b}). Recently, Knudsen and Hansen~\cite{knudsen_hansen_00} carried out numerical simulations avoiding the linearizations involved in Refs.~\cite{john_lubensky_85_86}, \cite{JLW_88} and \cite{roux_hansen_88}. They obtain $\varphi = 1.2$ in agreement with the series expansion result of Ref.~\cite{wang_lubensky_91}. 

It appears that a linearization of the network equations has a crucial effect in $d=2$. To our opinion this is plausible because the RJN is intimately related to the diluted $x$-$y$ model~\cite{foltin_stenull_janssen}. Hence one should expect that vortex excitations become important in $d=2$ and that a linearized description, corresponding to a spin wave approximation, is insufficient in this case. 

Last but not least in our little historical reminiscence we quote experimental values. Misra and Misiak measured $\varphi = 1.32$~\cite{misra_misiak_89a} and $\varphi = 1.45$~\cite{misra_misiak_89b}.

The previous paragraphs indicate that the status of diamagnetism in the RJN has several ramifications. On this basis it is hard to draw reliable conclusions on the nature of the diamagnetism in granular superconductors. Further work on this subject seems to be in order.

In this paper we determine the scaling behavior of $\chi^{(1)}$ and $\chi^{(2)}$ for transitions I and II by the powerful methods of renormalized field theory~\cite{amit_zinn-justin}. Our approach is based on the LGW type Hamiltonian introduced by JL. Using dimensional regularization and minimal subtraction we explore the renormalization of $\chi^{(1)}$ and $\chi^{(2)}$. Upon analyzing the general structure of the Feynman diagrams contributing to these renormalizations, we derive the critical behavior of $\chi^{(1)}$ and $\chi^{(2)}$ to arbitrary order in perturbation theory. 
The outline of this paper is as follows. In Sec.~\ref{model} we provide background on the model underlying our work. At first, we concretize the definition of the RJN and sketch its microscopic description. We mention the key physical quantities that are implicit in the microscopic model, such as the average tunneling current, and explain how they might be calculated by using the replica formalism. Then we condense the microscopic into a mesoscopic model that is represented by a field theoretic Hamiltonian $\mathcal{H}$. Our final expression for $\mathcal{H}$ is corroborated by a subsequent scaling analysis. In particular the irrelevance of a certain coupling associated with $T^2$ is revealed. Then we elaborate on several physical quantities that are native in the mesoscopic description. We give their definitions and explain how to extract them in the replica framework. A brief review of the RJN phase diagram concludes Sec.~\ref{model}. Section~\ref{renormalizationGroupAnalysis} contains the core of our renormalization group analysis. We gather the diagrammatic elements that are the ingredients of our perturbation calculation. Then we take a short detour and outline the renormalization and the scaling behavior of the order parameter correlation functions. This provides some background for our main task, the analysis of $\chi^{(1)}$ and $\chi^{(2)}$. Next, we calculate the Feynman diagrams contributing to  $\chi^{(1)}$ and $\chi^{(2)}$ at one-loop order. Equipped with some intuition about these diagrams we then determine their structural properties for arbitrary order in the loop expansion. At the end of Sec.~\ref{renormalizationGroupAnalysis} we describe the renormalization of $\chi^{(1)}$ and $\chi^{(2)}$, set up the corresponding renormalization group equations and determine the scaling behavior. In Sec.~\ref{concludingRemarks} we give a brief summary and concluding remarks. Technical details on the derivation of the full Gaussian propagator for the RJN can be found in appendix~\ref{app:propagator}.

\section{The model}
\label{model}

\subsection{The random Josephson network}
\label{rjn}
As mentioned briefly in the introduction, a RJN consists of superconducting grains located at the sites $i$ of a $d$-dimensional hypercubic lattice. Bonds between nearest neighboring sites are randomly occupied with probability $p$ by Josephson junctions and, respectively, empty with probability $1-p$~\cite{footnote1}. Each grain is characterized by a condensate wave function
\begin{eqnarray}
\label{waveFkt}
\Psi_i = \sqrt{\rho} \, \exp ( i \theta_i) \, .
\end{eqnarray}
Note, that the density of Cooper pairs on each grain is assumed to be a constant $\rho$ and that only the phase $\theta_i$  is allowed to fluctuate. The fixed amplitude approximation neglects charging effects due to quantum fluctuations~\cite{anderson_64}.  It is justified for grain sizes of the order of or smaller than both the bulk superconducting coherence length and the London penetration depth for the grains. 

The form of the wave function (\ref{waveFkt}) leads within the tight binding model to a quantummechanical expectation value for the total energy given by
\begin{eqnarray}
\label{quantumExpectation}
H = - \sum_{<i,j>} K_{i,j} \cos ( \delta_{i,j} ) \, .
\end{eqnarray}
$K_{i,j}$ is a hopping matrix element for the Cooper pairs. Here it is a random variable that takes on the value 1 with probability $p$ and the value 0 with probability $1-p$. The sum in Eq.~(\ref{quantumExpectation}) runs over all nearest neighbor pairs $<i,j>$. 
\begin{eqnarray}
\label{phaseDiff}
\delta_{i,j} = \theta_i - \theta_j - A_{i,j}
\end{eqnarray}
describes the phase difference between adjacent sites and
\begin{eqnarray}
A_{i,j} = e^\ast \int_i^j \brm{A} \cdot d \brm{l} \, .
\end{eqnarray}
Here $e^\ast$ is an abbreviation for $2\pi / \Phi_0$ with $\Phi_0$ being the flux quantum. The line integral is taken over an arbitrary differentiable curve from $i$ to $j$. $\brm{A}$ is the vector potential. We consider the gauge field $\brm{A}$ being entirely determined by a fixed external magnetic field and neglect fluctuations in $\brm{A}$. Expression (\ref{quantumExpectation}) governs the equilibrium statistical mechanics of the RJN and represents a Hamiltonian in the sense of statistical mechanics. Note the following important feature of $H$: it is invariant under the gauge transformation
\begin{eqnarray}
\label{gaugeTrafo}
\theta_i \to \theta_i + a (i)  \, ,\\
\brm{A} \to \brm{A} - \frac{1}{e^\ast} \nabla a \, ,
\end{eqnarray}
where $a$ is an arbitrary scalar function of the space coordinate.

A fundamental role in the RJN is played by the tunneling currents
\begin{eqnarray}
\label{tunnelCurrent}
I_{i,j} = - \frac{\partial H}{\partial A_{i,j}} = - K_{i,j} \sin ( \delta_{i,j} )  \, .
\end{eqnarray}
Their averages and correlation functions represent interesting observable quantities. Averaging over thermal degrees of freedom, henceforth indicated by $\langle \ldots \rangle_T$, may be discussed via the free energy
\begin{eqnarray}
\label{thermF}
F_T = - T \ln Z 
\end{eqnarray}
with the partition function $Z$ given by
\begin{eqnarray}
\label{partFunct}
Z = \int \mathcal{D} \theta \exp (-T^{-1} H) \, .
\end{eqnarray}
For convenience we have set the Boltzmann constant equal to one. $\int \mathcal{D} \theta$ is an abbreviation for $\int \prod_i d \theta_i$, where the product is taken over all lattice sites. 

Of course, a meaningful characterization of the statistical properties of the RJN requires more than just thermal averaging. In addition, a quenched average $[ \ldots ]_C$ over all possible configurations $C$ of the diluted network needs to be performed. This average can be achieved with help of the replica trick. $n$ copies of the network are considered simultaneously upon introducing the replicated Hamiltonian
\begin{eqnarray}
\label{repHamil} 
H (  \{  \vec{\delta} \} ) = \sum_{\alpha =1}^n  H (  \{  \delta^{(\alpha )} \} )   
= -  \sum_{\alpha =1}^n \sum_{<i,j>} K_{i,j} \cos ( \delta_{i,j}^{(\alpha )}) \, ,
\nonumber \\
\end{eqnarray}
where $\vec{\delta} = (\delta^{(1)}, \dots , \delta^{(n)})$. With this trick the configurationally averaged free energy
\begin{eqnarray}
\label{F} 
F = - T \left[  \ln Z \right]_C
\end{eqnarray}
can be written as
\begin{eqnarray}
\label{practicalF}
F =  \lim_{n \to 0} \frac{1}{n} F_n
\end{eqnarray}
with
\begin{eqnarray}
\label{practicalFn}
F_n = - T  \ln \left[  Z^n \right]_C \, .
\end{eqnarray}
The key benefit of this procedure is that the problem of averaging $\ln Z$ is basically replaced by the easier task of averaging $Z^n$. From the free energy various quantities of interest can be extracted upon taking derivatives. The average current is given by
\begin{eqnarray}
\left[ \left\langle I_{i,j} \right\rangle_T \right]_C = - \lim_{n \to 0} \frac{\partial F_n}{\partial A_{i,j}^{(\alpha )}} \, .
\end{eqnarray}
The derivation of correlation functions in the replica framework requires a little caution. Note that
\begin{eqnarray}
C_{i,j}^{(\alpha ,\beta )} = - \lim_{n \to 0} \frac{\partial^2 F_n}{\partial A_{i,j}^{(\alpha )} \partial A_{i,j}^{(\beta )}}
\end{eqnarray}
splits up into a replica diagonal and a replica independent part. Due to the permutation symmetry between the replicas $C_{i,j}^{(\alpha ,\beta )}$ is of the form
\begin{eqnarray}
C_{i,j}^{(\alpha ,\beta )} = C_{i,j}^{(1)} \delta_{\alpha ,\beta} + C_{i,j}^{(2)} ,
\end{eqnarray}
with
\begin{eqnarray}
\label{defC1}
C_{i,j}^{(1)} = T^{-1} \left\{ \left[ \left\langle I_{i,j}^2 \right\rangle_T \right]_C - [ \left\langle I_{i,j} \right\rangle_T^2 ]_C \right\} - E_c \end{eqnarray}
and
\begin{eqnarray}
\label{defC2}
C_{i,j}^{(2)} = T^{-1} \left\{ [ \left\langle I_{i,j} \right\rangle_T^2 ]_C - \left[ \left\langle I_{i,j} \right\rangle_T \right]_C^2 \right\} \, .
\end{eqnarray}
The $E_c$ emerging in Eq.~(\ref{defC1}) stands for the condensation energy
\begin{eqnarray}
\label{defEc}
E_c =  \left[ \left\langle K_{i,j} \cos ( \delta_{i,j} ) \right\rangle_T \right]_C  \, .
\end{eqnarray}

\subsection{Field theoretic Hamiltonian}
\label{fieldTheoreticHamiltonian}
Now we proceed towards a field theoretic model for the RJN. The following derivation of a field theoretic Hamiltonian is guided by the work of JL.

Our starting point is the observation, that the $[Z^n]_C$ appearing in Eq.~(\ref{practicalF}) can be written as
\begin{eqnarray}
[Z^n]_C = \int \mathcal{D} \vec{\theta} \, \exp (-T^{-1} H_{\text{eff}}) 
\end{eqnarray}
with an effective Hamiltonian
\begin{eqnarray}
H_{\text{eff}} = - \ln \left[  \exp \left[  -T^{-1} H \left(  \left\{  \vec{\delta} \right\} \right) \right] \right]_C \, .
\end{eqnarray}
By virtue of the replica approach we can perform the quenched average once and for all at this early stage. This leads to
\begin{eqnarray}
\label{effHamil}
H_{\text{eff}} = \sum_{<i,j>} K \left(  \vec{\delta}_{i,j} \right) \, ,
\end{eqnarray}
with
\begin{eqnarray}
\label{combo}
K \left(  \vec{\delta} \right) = - \ln \left[  1 + \upsilon \exp \left( T^{-1} \sum_{\alpha =1}^n \cos \left(  \delta^{(\alpha )}\right) \right) \right]   \, ,
\nonumber \\
\end{eqnarray}
where $\upsilon = p/(1-p)$. In Eq.~(\ref{effHamil}) we have dropped a constant term $N_B \ln (1-p)$, where $N_B$ stands for the number of bonds in the undiluted lattice.

Next, we adopt ideas developed by Stephen~\cite{stephen_78} in the context of the RRN. The idea here is to introduce the quantity
\begin{eqnarray}
\label{defPsi}
\psi_{\vec{\lambda}}(i) = \exp \left( i \vec{\lambda} \cdot \vec{\theta}_i \right) \, , \quad \vec{\lambda} \neq \vec{0} \, .
\end{eqnarray}
As we go along, this quantity will grow into the role of an order parameter field. The $\vec{\lambda}$ appearing in (\ref{defPsi}) is an $n$-component vector in replica space, $\vec{\lambda} = ( \lambda^{(1)}, \ldots , \lambda^{(n)} )$. The dot product in Eq.~(\ref{defPsi}) is defined as $\vec{\lambda} \cdot \vec{\theta} = \sum_{\alpha =1}^n \lambda^{(\alpha)} \theta^{(\alpha)}$. The condition $\vec{\lambda} \neq \vec{0}$ is imposed in order to qualify $\psi_{\vec{\lambda}}(i)$ as an order parameter. That is simply because $\psi_{\vec{0}}(i)$ is equal to one and hence, being a constant, not capable of sensing a phase transition. The components $\lambda^{(\alpha)}$ are chosen to take on integral values. With this choice, the $\exp ( i \vec{\lambda} \cdot \vec{\theta} )$ represent a complete set of orthonormal functions satisfying the orthonormality and completeness relations
\begin{subequations} 
\label{orthoComplete}
\begin{eqnarray}
\label{ortho}
\frac{1}{(2\pi )^n} \int_{-\pi}^\pi d^n \theta \, \exp \left( i \vec{\lambda} \cdot \vec{\theta} \right)  = \delta_{\vec{\lambda}, \vec{0}}
\end{eqnarray}
and
\begin{eqnarray}
\label{complete}
\frac{1}{(2\pi )^n} \sum_{\vec{\lambda}}  \exp \left( - i \vec{\lambda} \cdot \vec{\theta} \right)  =  \delta \left( \vec{\theta} \right) \, .
\end{eqnarray}
\end{subequations}
Based on these relations one can define the replica space Fourier transform
\begin{eqnarray}
\phi \left(  \vec{\theta} , i \right) =  \sum_{\vec{\lambda} \neq \vec{0}}   \exp \left( - i \vec{\lambda} \cdot \vec{\theta} \right) \psi_{\vec{\lambda}}(i) 
\end{eqnarray}
of $\psi_{\vec{\lambda}}(i)$. Note that the condition $\vec{\lambda} \neq \vec{0}$ on $\psi_{\vec{\lambda}}(i)$ transforms into the condition
\begin{eqnarray}
\frac{1}{(2\pi )^n} \int_{-\pi}^\pi d^n \theta \,   \phi \left(  \vec{\theta} , i \right) = 0 
\end{eqnarray}
on $\phi (  \vec{\theta} , i )$. Thus, $\phi (  \vec{\theta} , i )$ can be interpreted as a (continuously indexed) Potts spin~\cite{Zia_Wallace_75} .

Now we expand $K (  \vec{\delta}_{i,j} )$ in terms of $\psi_{\vec{\lambda}}(i)$. Rewriting $K (  \vec{\delta}_{i,j} )$ as
\begin{eqnarray}
K \left(  \vec{\delta}_{i,j} \right) =  \int_{-\pi}^\pi d^n \delta \, K \left(  \vec{\delta} \right) \delta \left( \vec{\delta} -  \vec{\delta}_{i,j} \right)  
\end{eqnarray}
we obtain after a little algebra
\begin{eqnarray}
\label{hunger}
K \left(  \vec{\delta}_{i,j} \right) =   \sum_{\vec{\lambda} \neq \vec{0}}  \psi_{\vec{\lambda}}(i) \psi_{-\vec{\lambda}}(j) \exp \left( - i \vec{\lambda} \cdot \vec{A}_{i,j}  \right) \widetilde{K} \left( \vec{\lambda} \right) \, .
\nonumber \\
\end{eqnarray}
$\widetilde{K} ( \vec{\lambda} )$ is defined as the replica space Fourier transform of $K (  \vec{\delta} )$:
\begin{eqnarray}
\label{FT}
\widetilde{K} \left( \vec{\lambda} \right) = \frac{1}{(2\pi )^n} \int_{-\pi}^\pi d^n \delta  \,  \exp \left( i \vec{\lambda} \cdot \vec{\delta} \right)   K \left(  \vec{\delta} \right) \, .
\end{eqnarray}
In writing Eq.~(\ref{hunger}) we have dropped a constant term $\widetilde{K} (  \vec{0} )$. To evaluate Eq.~(\ref{FT}) further, we insert
(\ref{combo}) and expand the logarithm. We arrive at
\begin{eqnarray}
\widetilde{K} \left( \vec{\lambda} \right) = \sum_{l=1}^\infty \frac{(-1)^{l}}{l} \, \upsilon^l F_l \left( \vec{\lambda} \right)
\end{eqnarray}
with
\begin{eqnarray}
\label{efel}
&& F_l \left( \vec{\lambda} \right) = \frac{1}{(2\pi )^n}  \int_{-\pi}^\pi d^n \delta  \,  
\nonumber \\
&& \times \, \exp \left[  \sum_{\alpha =1}^n \left[ i \lambda^{(\alpha )}  \delta^{(\alpha )}  + l T^{-1} \cos \left(  \delta^{(\alpha )}  \right) \right] \right] \, .
\end{eqnarray}
The integral in Eq.~(\ref{efel}) can be evaluated in the low temperature limit by employing the saddle point method. Note that this step amounts to linearization of the network equations, or in other words, to a spin wave approximation. For $n\to 0$ this procedure leads to
\begin{eqnarray}
\label{expK}
\widetilde{K} \left( \vec{\lambda} \right) = \tau + w \vec{\lambda}^2 + O \left(  T^2 \right) \, ,
\end{eqnarray}
where $\tau = \tau (p)$ and $w = w(p,T)\sim T$ are expansion coefficients. In Sec.~\ref{relevanceNote} we will show explicitly, that higher order terms in (\ref{expK}) lead to irrelevant contributions in the field theoretic formulation, and that, hence, their omission is justified.

Collecting we find the following expression for the effective Hamiltonian:
\begin{eqnarray}
\label{effHamilWeit}
&& H_{\text{eff}} = \sum_{<i,j>} \sum_{\vec{\lambda} \neq \vec{0}}  \psi_{\vec{\lambda}}(i) \psi_{-\vec{\lambda}}(j)
\nonumber \\
&& \times \,  \exp \left( - i \vec{\lambda} \cdot \vec{A}_{i,j}  \right)  \left\{  \tau + w \vec{\lambda}^2 + \ldots \right\} \, .
\end{eqnarray}
At this stage we carry out a gradient expansion. We have to pay regard to the fact that the effective Hamiltonian  as given in Eq.~(\ref{effHamilWeit}) is invariant under the gauge transformation
\begin{subequations} 
\label{gaugeTrafo2}
\begin{eqnarray}
\label{gt2a}
\psi_{\vec{\lambda}}(i) &\to& \psi_{\vec{\lambda}}(i) \exp \left[   i \vec{\lambda} \cdot \vec{a} (i) \right] \, ,
\\
\label{gt2b}
\vec{\brm{A}} (i) &\to&  \vec{\brm{A}} (i) - \frac{1}{e^\ast} \nabla \vec{a} (i) \, .
\end{eqnarray}
\end{subequations}
Hence, we must keep only those terms in the gradient expansion that comply with this invariance. We obtain
\begin{eqnarray}
\label{effHamilWeiter}
&&H_{\text{eff}} = \frac{1}{2} \sum_{i} \sum_{\underline{b}_i} \sum_{\vec{\lambda} \neq \vec{0}}  \psi_{\vec{\lambda}}(i) 
\nonumber \\
&& \times \,
\Big\{  1 + \frac{1}{2} \Big[ \underline{b}_i \cdot \left( \nabla - i e^\ast \vec{\lambda} \cdot \vec{\brm{A}} (i) \right) \Big]^2 \Big\}
\psi_{-\vec{\lambda}}(i)
\nonumber \\
&& \times \,  \left\{  \tau + w \vec{\lambda}^2 + \ldots \right\} \, ,
\end{eqnarray}
where the second sum runs over all lattice vectors $\underline{b}_i$ between site $i$ and its nearest neighbors. 

We proceed with the usual coarse graining step and replace $\psi_{\vec{\lambda}}(i)$ by the order parameter field $\psi_{\vec{\lambda}}(\brm{x})$. The order parameter field inherits the constraint $\vec{\lambda} \neq \vec{0}$. Then, a mesoscopic free energy in the spirit of Landau is devised. Guided by Eq.~(\ref{effHamilWeiter}) we write down the Landau-Ginzburg-Wilson type Hamiltonian
\begin{eqnarray}
\label{hamiltonian}
&&{\mathcal{H}} = \int d^dx \bigg\{ \frac{1}{2} \sum_{\vec{\lambda} \neq \vec{0}} \psi_{-\vec{\lambda}} \left( {\rm{\bf x}} \right)  K \left(  \nabla , \vec{\lambda} , \vec{\brm{A}} \right) \psi_{\vec{\lambda}} \left( {\rm{\bf x}} \right)
\nonumber \\
&& + \, \frac{g}{6} \sum_{\vec{\lambda}, \vec{\lambda}^\prime  , \vec{\lambda} + \vec{\lambda}^\prime \neq \vec{0}} \psi_{-\vec{\lambda}} \left( {\rm{\bf x}} \right) \psi_{-\vec{\lambda}^\prime} \left( {\rm{\bf x}} \right) \psi_{\vec{\lambda} + \vec{\lambda}^\prime} \left( {\rm{\bf x}} \right)  \bigg\} \, ,
\end{eqnarray}
 where terms of higher order in the fields have been neglected since they turn out to be irrelevant. The kernel appearing in the Eq.~(\ref{hamiltonian}) is given by
\begin{eqnarray}
\label{kernel}
K \left(  \nabla , \vec{\lambda} , \vec{\brm{A}} \right) = \tau + w \vec{\lambda}^2 - \Big[ \nabla - i \vec{\lambda} \cdot \vec{\brm{A}} \Big]^2 \, .
\end{eqnarray}
The coefficients $\tau$ and $w$ should be understood as the coarse grained analogs of the original coefficients featured in Eq.~(\ref{effHamilWeiter}). Similarly, the $\vec{\brm{A}}$ in the kernel (\ref{kernel}) is a coarse grained version of the original gauge field. The coarse grained $\vec{\brm{A}}$ is defined so that it incorporates the charge $e^\ast$.

It must be emphasized that $\mathcal{H}$ is invariant under the gauge transformation
\begin{subequations} 
\label{gaugeTrafo3}
\begin{eqnarray}
\label{gt3a}
\psi_{\vec{\lambda}}( \brm{x}) &\to& \psi_{\vec{\lambda}}(\brm{x}) \exp \left[   i \vec{\lambda} \cdot \vec{a} (\brm{x}) \right] \, ,
\\
\label{gt3b}
\vec{\brm{A}} (\brm{x}) &\to&  \vec{\brm{A}} (\brm{x}) - \nabla \vec{a} (\brm{x}) \, ,
\end{eqnarray}
\end{subequations}
with the components of $\vec{a}$ being arbitrary scalar functions of $\brm{x}$. This gauge invariance will have important consequences as we go along.

We point out that $\mathcal{H}$ resembles for vanishing $\vec{\brm{A}}$ the form of the field theoretic Hamiltonian for the RRN as studied by Harris and Lubensky~\cite{harris_lubensky_87a} and the present authors~\cite{stenull_janssen_oerding_99,stenull_2000}. For vanishing $\vec{\brm{A}}$, the only formal distinction resides in the different domains of $\vec{\lambda}$. In the limit $w \to 0$~\cite{footnote2}, however, this difference has no consequence and the perturbation expansions for the RJN and the RRN coincide. For $w=0$, in particular, both models reduce to purely geometric percolation.

\subsection{A note on relevance}
\label{relevanceNote}
Here we will show that it is indeed justified to truncate the expansion~(\ref{expK}) at first order in $T$. In other words, we will show that the higher order terms are irrelevant in the sense of the renormalization group. Our actual tool will be a scaling analysis in the replica variable $\vec{\lambda}$.

Now suppose we had retained higher order terms in the expansion~(\ref{expK}). Then the kernel of $\mathcal{H}$ would be of the form
\begin{eqnarray}
\label{kernelLong}
K \left(  \nabla , \vec{\lambda} , \vec{\brm{A}} \right) = \tau + w \vec{\lambda}^2 + \sum_{k=2}^\infty  w_k \vec{\lambda}^{2k} - \Big[ \nabla - i \vec{\lambda} \cdot \vec{\brm{A}} \Big]^2 
\nonumber \\
\end{eqnarray}
with $w_k \sim T^k$. We facilitate our scaling analysis by setting $\vec{\lambda} \to b^{-1} \vec{\lambda}$, where $b$ is some scaling factor. Upon substituting $\psi_{\vec{\lambda}} ( {\rm{\bf x}} ) = \psi_{b^{-1} \vec{\lambda}}^\dag ( {\rm{\bf x}} )$ into the Hamiltonian we get
\begin{eqnarray}
\label{scaling1}
&&\mathcal{H} \left[ \psi_{b^{-1} \vec{\lambda}}^\dag ( {\rm{\bf x}} ) ,  \vec{\brm{A}} (\brm{x}) ;\tau, w, \{ w_k \} \right] 
\nonumber \\
&& = \, \int d^dx \, \Bigg\{ \frac{1}{2} \sum_{\vec{\lambda} \neq \vec{0} } \psi_{b^{-1} \vec{\lambda}}^\dag ( {\rm{\bf x}} ) \, K \left( \nabla 
,\vec{\lambda}, \vec{\brm{A}} \right) \psi_{- b^{-1} \vec{\lambda}}^\dag ( {\rm{\bf x}} )
\nonumber \\
&& + \, 
\frac{g}{6} \sum_{\vec{\lambda}, \vec{\lambda}^\prime  , \vec{\lambda} + \vec{\lambda}^\prime \neq \vec{0}} \psi_{-b^{-1}\vec{\lambda}}^\dag \left( {\rm{\bf x}} \right) \psi_{-b^{-1}\vec{\lambda}^\prime}^\dag \left( {\rm{\bf x}} \right) 
\nonumber \\
&& \times \,
\psi_{b^{-1}\vec{\lambda} + b^{-1}\vec{\lambda}^\prime}^\dag \left( {\rm{\bf x}} \right) \Bigg\} \, .
\end{eqnarray}
Renaming the scaled replica variables $\vec{\lambda}^\dag = b^{-1} \vec{\lambda}$ leads to
\begin{eqnarray}
\label{scaling2}
&&\mathcal{H} \left[ \psi_{\vec{\lambda}^\dag}^\dag ( {\rm{\bf x}} ) ,  \vec{\brm{A}} (\brm{x}) ; \tau, w, \{ w_k \}  \right] 
\nonumber \\
&& = \, \int d^dx \, \Bigg\{ \frac{1}{2} \sum_{\vec{\lambda} \neq \vec{0} } \psi_{\vec{\lambda}^\dag}^\dag ( {\rm{\bf x}} ) \, K \left( \nabla , b \vec{\lambda}^\dag , \vec{\brm{A}} \right) \psi_{-\vec{\lambda}^\dag}^\dag ( {\rm{\bf x}} )
\nonumber \\
&& + \, 
\frac{g}{6} \sum_{\vec{\lambda}, \vec{\lambda}^\prime  , \vec{\lambda} + \vec{\lambda}^\prime \neq \vec{0}} \psi_{-\vec{\lambda}^\dag}^\dag \left( {\rm{\bf x}} \right) \psi_{-\vec{\lambda}^{\prime ^\dag}}^\dag \left( {\rm{\bf x}} \right) 
\psi_{\vec{\lambda}^\dag +\vec{\lambda}^{\prime ^\dag}}^\dag \left( {\rm{\bf x}} \right) \Bigg\} \, .
\nonumber \\
\end{eqnarray}
Now we are going to exploit an important feature of the summations over the replica variable $\vec{\lambda}$. In the low temperature limit, i.e., for $w \to 0$,  the summation $\sum_{\vec{\lambda} \neq \vec{0}} \ldots$ can be replaced by the integration $\int_{-\infty}^\infty d^n \lambda \, \ldots$. Poisson's summation formula guarantees that the neglected terms are of the order $\exp ( - \text{const}/w)$~\cite{footnote:reminder}. In the continuum formulation the rescaling leads to $b^n \int_{-\infty}^\infty d^n \lambda^\dag \, \ldots$. Hence, the scaling factor $b$ drops out in the limit $n\to 0$ and we can identify $\vec{\lambda}^\dag$ with $\vec{\lambda}$. We are led to the conclusion
\begin{eqnarray}
\label{relForH}
&&\mathcal{H} \left[ \psi_{b^{-1} \vec{\lambda}} ( {\rm{\bf x}} ) ,  \vec{\brm{A}} (\brm{x}) ; \tau, w, \{ w_k \} \right] 
\nonumber \\
&&=\, \mathcal{H} \left[ \psi_{\vec{\lambda}} ( {\rm{\bf x}} ) , b \vec{\brm{A}} (\brm{x}) ; \tau, b^2 w, \{ b^{2k} w_k \}  \right] \, . 
\end{eqnarray}

Next we consider the implications of Eq.~(\ref{relForH}) on the free energy. In the present field theoretic formulation, the Helmholts free energy is defined as
\begin{eqnarray}
\label{correl}
&&\mathcal{F} \left[ \vec{\brm{A}} (\brm{x}); T, \tau , w, \{ w_k \}  \right]   = -T  \ln \mathcal{Z}
\end{eqnarray}
with the partition function
\begin{eqnarray}
\label{defFieldZ}
\mathcal{Z} = \int \mathcal{D} \psi \,  \exp \left( - T^{-1}\mathcal{H} \left[ \psi_{\vec{\lambda}} ( {\rm{\bf x}} ) , \vec{\brm{A}} (\brm{x}) ; \tau, w, \{  w_k \}  \right] \right) \, .
\nonumber \\
\end{eqnarray}
Here, $\mathcal{D} \psi$ indicates an integration over the set of variables $\{ \psi_{\vec{\lambda}} ( {\rm{\bf x}} ) \}$ for all ${\rm{\bf x}}$ and $\vec{\lambda}$. Equation~(\ref{relForH}) implies that
\begin{eqnarray}
\label{relF}
&&\mathcal{F} \left[ \vec{\brm{A}} (\brm{x}); T, \tau , w, \{ w_k \}  \right] 
\nonumber \\
&&= \, \mathcal{F} \left[ b \vec{\brm{A}} (\brm{x}); T, \tau , b^2 w, \{  b^{2k} w_k \}  \right]  \, .
\end{eqnarray}
Of course, we are free to choose the scaling parameter to our liking. With the choice $b^2 = w^{-1}$ we obtain
\begin{eqnarray}
\label{relF2}
&&\mathcal{F} \left[ \vec{\brm{A}} (\brm{x}); T, \tau , w, \{ w_k \}  \right] 
\nonumber \\
&&= \, \mathcal{F} \left[ w^{-1/2} \vec{\brm{A}} (\brm{x});T,  \tau , 1, \left\{   \frac{w_k}{w^k} \right\}  \right]  \, .
\end{eqnarray}
We learn from Eq.~(\ref{relF2}) that the coupling constants $w_k$ appear only in the combination $w_k / w^k$. A trivial consequence of the fact that the Hamiltonian $\mathcal{H}$ must be dimensionless is that $w \vec{\lambda}^2 \sim \mu^2$ and  $w_k \vec{\lambda}^{2k} \sim \mu^2$, where $\mu$ is an inverse length scale. In other words, $w \vec{\lambda}^2$ and $w_k \vec{\lambda}^{2k}$ have a naive dimension 2. Thus, $w_k / w^k \sim \mu^{2-2k}$ and hence the $w_k / w^k$ have a negative naive dimension. This leads to the conclusion that the $w_{k}/ w^k$ are irrelevant couplings and that the leading critical behavior of the free energy is described by
\begin{eqnarray}
\label{relF3}
\mathcal{F} \left[ \vec{\brm{A}} (\brm{x}); T, \tau , w \right] = f \left[ w^{-1/2} \vec{\brm{A}} (\brm{x}); T, \tau \right] \, ,
\end{eqnarray}
where $f$ is some functional of $w^{-1/2} \vec{\brm{A}} (\brm{x})$.

The fact that $w_2$ appears only in an irrelevant combination was overlooked in Ref.~\cite{JLW_88}. This ultimatively led to an erroneous prediction for the scaling behavior of $\chi^{(2)}$. 

\subsection{Current density, magnetization and related quantities}
In this section we elaborate on various physical quantities embedded in the field theoretic model. We provide, within the replica framework, definitions of the current density and the magnetization along with their averages and correlation functions. We explain the physical content of replica quantities and describe how it can be extracted.

The role occupied in the original microscopic model by the replicated tunneling currents $I_{i,j}^{(\alpha )}$ is taken in the field theoretic formulation by the replicated current density
\begin{eqnarray}
\label{defJ}
J_i^{(\alpha )} (\brm{x}) = - \frac{\delta \mathcal{H}}{\delta A_i^{(\alpha )} (\brm{x})} \, .
\end{eqnarray}
Note that the index $i$ specifies here the component of the current density in $d$-dimensional space and should not be confused with the site $i$. 

The current density has a very important feature. It represents the Noether current associated with the gauge invariance of the Hamiltonian $\mathcal{H}$. Hence, it satisfies the conservation relation
\begin{eqnarray}
\label{manifest}
\nabla \cdot \brm{J}^{(\alpha )} (\brm{x}) = 0 \, .
\end{eqnarray}

Averages in the field theoretic formulation are declared by means of the functional integral
\begin{eqnarray}
\langle \cdots \rangle = \frac{1}{\mathcal{Z}} \int \mathcal{D} \psi \cdots \exp \left(  - T^{-1}\mathcal{H} \right)  \, .
\end{eqnarray}
Various of these averages can be extracted from the free energy introduced in Sec.~\ref{relevanceNote}.   This free energy can be expanded as
\begin{eqnarray}
&&\mathcal{F} \big[  \vec{\brm{A}} \big] = \mathcal{F} \big[  \vec{\brm{0}}  \big]
-  \int d^d x \, \left\langle  J_i^{(\alpha )} (\brm{x}) \right\rangle A_i^{(\alpha )} (\brm{x})
\nonumber \\
&& - \, \frac{1}{2} \int d^d x \int d^d x ^\prime \, C_{i,j}^{(\alpha ,\beta)} (\brm{x} - \brm{x}^\prime)  A_i^{(\alpha )} (\brm{x}) A_j^{(\beta )} (\brm{x}^\prime) 
\nonumber \\
&& + \, \ldots \, ,
\end{eqnarray}
where the summation convention is understood for the indices labeling space and replica coordinates.
\begin{eqnarray}
\left\langle J_i^{(\alpha )} (\brm{x}) \right\rangle = - \left. \frac{\delta \mathcal{F}}{\delta A_i^{(\alpha )} (\brm{x})} \right|_{\vec{\brm{A}} = \vec{\brm{0}}}
\end{eqnarray}
is the average replica current density. The second order term features the correlation function
\begin{eqnarray}
\label{currCorr}
&& C_{i,j}^{(\alpha ,\beta)} (\brm{x} - \brm{x}^\prime) = - \left. \frac{\delta^2 \mathcal{F}}{\delta A_i^{(\alpha )} (\brm{x}) \delta A_j^{(\beta )} (\brm{x}^\prime)} \right|_{\vec{\brm{A}} = \vec{\brm{0}}}
\nonumber \\
&& = \,T^{-1} \Big\{ \left\langle J_i^{(\alpha )} (\brm{x}) J_j^{(\beta )} (\brm{x}^\prime ) \right\rangle - \left\langle J_i^{(\alpha )} (\brm{x}) \right\rangle \left\langle J_j^{(\beta )} (\brm{x}^\prime ) \right\rangle \Big\}
\nonumber \\
&& - \, \delta (\brm{x} - \brm{x}^\prime ) \delta_{i,j} \sum_{\vec{\lambda} \neq \vec{0}} \lambda^{(\alpha )} \lambda^{(\beta )} \left\langle \psi_{\vec{\lambda}} (\brm{x} ) \psi_{-\vec{\lambda}} (\brm{x}^\prime ) \right\rangle \, .
\end{eqnarray}
The relation between the field theoretic average of the replica current density and the thermal and configurational average of the physical current density is straightforward:
\begin{eqnarray}
\lim_{n \to 0}\left\langle J_i^{(\alpha )} (\brm{x}) \right\rangle = \left[  \left\langle J_i (\brm{x}) \right\rangle \right]_C \, .
\end{eqnarray}
As far as correlation functions are concerned, the situation is somewhat more subtle. The structure of the correlation functions must be so that they are invariant under permutations of the replicas. Hence the two-point functions are of the form
\begin{eqnarray}
C_{i,j}^{(\alpha ,\beta)} (\brm{x} - \brm{x}^\prime) = C_{i,j}^{(1)} (\brm{x} - \brm{x}^\prime) \,  \delta_{\alpha ,\beta} + C_{i,j}^{(2)} (\brm{x} - \brm{x}^\prime)
\nonumber \\
\end{eqnarray}
with
\begin{eqnarray}
\label{formC1}
&&C_{i,j}^{(1)} (\brm{x} - \brm{x}^\prime) = T^{-1} \big\{ \left[ \left\langle J_i (\brm{x}) J_j (\brm{x}^\prime ) \right\rangle_T \right]_C 
\nonumber \\
&&-\,  \left[ \left\langle J_i (\brm{x}) \right\rangle_T  \left\langle J_j (\brm{x}^\prime ) \right\rangle_T \right]_C \big\}- \delta_{i,j} \delta (\brm{x} - \brm{x}^\prime ) E^{(1)}_c
\nonumber \\
\end{eqnarray}
and
\begin{eqnarray}
&&C_{i,j}^{(2)} (\brm{x} - \brm{x}^\prime) = T^{-1} \big\{ \left[ \left\langle J_i (\brm{x}) \right\rangle_T  \left\langle J_j (\brm{x}^\prime ) \right\rangle_T \right]_C
\nonumber \\
&&-\,  \left[ \left\langle J_i (\brm{x}) \right\rangle_T   \right]_C \left[ \left\langle J_j (\brm{x}^\prime ) \right\rangle_T \right]_C \big\} - \delta_{i,j} \delta (\brm{x} - \brm{x}^\prime ) E^{(2)}_c \, ,
\nonumber \\
\end{eqnarray}
where the replica limit $n \to 0$ is understood. $E^{(1)}_c$ and $E^{(2)}_c$ are the replica diagonal and the replica independent part of
\begin{eqnarray}
E^{(\alpha ,\beta)}_c = \sum_{\vec{\lambda} \neq \vec{0}} \lambda^{(\alpha )} \lambda^{(\beta )} \left\langle \psi_{\vec{\lambda}} (\brm{x} ) \psi_{-\vec{\lambda}} (\brm{x}^\prime ) \right\rangle \, .
\end{eqnarray}

From Eq.~(\ref{formC1}) one learns that
\begin{eqnarray}
C_{i,j}^{(1)} (\brm{x} - \brm{x}^\prime) = \left[  C_{i,j} (\brm{x}, \brm{x}^\prime) \right]_C
\end{eqnarray}
for $n\to 0$, i.e., the replica diagonal part of $C_{i,j}^{(\alpha ,\beta)} (\brm{x}, \brm{x}^\prime)$ corresponds to the average of the physical density current correlation functions $C_{i,j} (\brm{x}, \brm{x}^\prime)$ over all configurations $C$. The physical content of $C_{i,j}^{(2)}$ will become clear below.

Now we shift focus and turn from the current densities to the magnetization and its correlations. An external magnetic field can be introduced into the model via
\begin{eqnarray}
F_{i,j} (\brm{x}) = \partial_i A_j (\brm{x}) - \partial_j A_i (\brm{x}) \, .
\end{eqnarray}
For $d=3$ this reduces to the usual $\brm {B} = \text{rot} \brm{A}$ with the components of the magnetic field given by $B_3 = F_{1,2}$ and so on. Having $F_{i,j}$ at hand, we can rewrite the expansion of the free energy as
\begin{eqnarray}
\label{neueExp}
&&\mathcal{F} \big[  \vec{\brm{A}} \big] = \mathcal{F} \big[  \vec{\brm{0}}  \big]
-  \int d^d x \, \left\langle  J_i^{(\alpha )} (\brm{x}) \right\rangle A_i^{(\alpha )} (\brm{x})
\nonumber \\
&& - \, \frac{1}{4} \int d^d x \int d^d x ^\prime \, C^{(\alpha ,\beta)} (\brm{x} - \brm{x}^\prime) F_{i,j}^{(\alpha )} (\brm{x}) F_{i,j}^{(\beta )} (\brm{x}^\prime) 
\nonumber \\
&& \, + \ldots \, .
\end{eqnarray}
In recasting the second order term we have exploited that
\begin{eqnarray}
\label{corro}
\partial_i C_{i,j}^{(\alpha ,\beta)} (\brm{x}) = 0
\end{eqnarray}
by virtue of the gauge invariance and its manifestation (\ref{manifest}). Due to Eq.~(\ref{corro}), the Fourier transform
\begin{eqnarray}
\widetilde{C}_{i,j}^{(\alpha ,\beta)} (\brm{k}) = \int d^d x \, C_{i,j}^{(\alpha ,\beta)} (\brm{x}) \exp \left( - i  \brm{k} \cdot \brm{x} \right)
\end{eqnarray}
of $C_{i,j}^{(\alpha ,\beta)} (\brm{x})$ is of the form
\begin{eqnarray}
\label{corroFourier}
\widetilde{C}_{i,j}^{(\alpha ,\beta)} (\brm{k}) = \left( \brm{k}^2 \delta_{i,j} - k_i k_j \right) \widetilde{C}^{(\alpha ,\beta)} (\brm{k}) \, .
\end{eqnarray}
The $C^{(\alpha ,\beta)} (\brm{x})$ in the expansion~(\ref{neueExp}) is defined as nothing but the Fourier transform of $\widetilde{C}^{(\alpha ,\beta)} (\brm{k})$. 

In the remainder of this paper we will be concerned with a homogeneous external magnetic field perpendicular to the 1-2-plane ($x$-$y$-plane). In three dimensions this corresponds to a $B$ field pointing in the 3-direction ($z$-direction). We formulate the envisaged magnetic field by setting
\begin{eqnarray}
A_i (\brm{x}) = \frac{1}{2} x_k F_{k,i} 
\end{eqnarray}
with
\begin{eqnarray}
F_{k,i} = B \left(  \delta_{k,1} \delta_{i,2} - \delta_{k,2} \delta_{i,1} \right) \, .
\end{eqnarray}
With help of this expression we obtain the free energy as a function of the replicated field amplitude $\vec{B}$:
\begin{eqnarray}
\label{neueExpB}
&&\mathcal{F} \big(  \vec{B} \big) = \mathcal{F} \big(  \vec{0}  \big)
-  \int d^d x \, M^{(\alpha )} (\brm{x}) \, B^{(\alpha )} 
\nonumber \\
&& - \, \frac{1}{2} \int d^d x \int d^d x ^\prime \, \chi^{(\alpha ,\beta)} (\brm{x} - \brm{x}^\prime) \, B^{(\alpha )}  B^{(\beta )}
\nonumber \\
&& \, + \ldots \, ,
\end{eqnarray}
where
\begin{eqnarray}
\label{relMag}
M^{(\alpha )} (\brm{x}) = \frac{1}{2} \left[ \left\langle  J_2^{(\alpha )} (\brm{x}) \right\rangle x_1 - \left\langle  J_1^{(\alpha )} (\brm{x}) \right\rangle x_2 \right]
\end{eqnarray}
and
\begin{eqnarray}
\label{relChi}
\chi^{(\alpha ,\beta)} (\brm{x}- \brm{x}^\prime ) =  C^{(\alpha ,\beta)} (\brm{x} - \brm{x}^\prime) \, .
\end{eqnarray}
Adapting the usual definition of the magnetization in the homogeneous field setup to the replica framework we have 
\begin{eqnarray}
M^{(\alpha )} = - \frac{1}{V} \left. \frac{\partial \mathcal{F}}{\partial B^{(\alpha )} } \right|_{\vec{B} = \vec{0}} =  \frac{1}{V} \, M^{(\alpha )}_{\text{tot}} \, ,
\end{eqnarray}
where $V$ stands for the volume of the system and $M^{(\alpha )}_{\text{tot}}$ is the replica version of the total magnetic moment
\begin{eqnarray}
M_{\text{tot}} =  \int d^d x \, M (\brm{x}) \, .
\end{eqnarray}
The physical magnetization $M$ is retrieved by taking the limit $n\to 0$. From (\ref{relMag}) one obtains immediately that $M$ vanishes for $B=0$. Turning to the diamagnetic susceptibility we have
\begin{eqnarray}
\label{replicaChi}
&&\chi^{(\alpha ,\beta)} = - \frac{1}{V} \left. \frac{\partial^2 \mathcal{F}}{\partial B^{(\alpha )} \partial B^{(\beta)}} \right|_{\vec{B} = \vec{0}} 
\nonumber \\
&&=  \, \frac{1}{V} \, \int d^d x \int d^d x ^\prime \, \chi^{(\alpha ,\beta)} (\brm{x} - \brm{x}^\prime) \, .
\end{eqnarray}
A glance at (\ref{relChi}) brings about two important observations. First, we see that
\begin{eqnarray}
\label{relChiC}
\chi^{(\alpha ,\beta)} =  \widetilde{C}^{(\alpha ,\beta)} (\brm{0})  \, .
\end{eqnarray}
This relation will play an important role in our actual calculations. The second observation is that the diamagnetic susceptibility should have the same replica structure as the current density correlations, i.e., 
\begin{eqnarray}
\label{propChi}
\chi^{(\alpha ,\beta)} =  \chi^{(1)} \, \delta_{\alpha ,\beta} +   \chi^{(2)}  \, .
\end{eqnarray}
In the replica limit $\chi^{(1)}$ has the physical content
\begin{eqnarray}
\chi^{(1)} = \left[ \chi \right]_C  
\end{eqnarray}
with $\chi$ being the diamagnetic susceptibility for a given configuration $C$. $\chi^{(2)}$ contains the fluctuations of the total magnetization. For $n\to 0$ one has
\begin{eqnarray}
&&\chi^{(2)} = \frac{1}{V} \, \int d^d x \int d^d x ^\prime 
\nonumber \\
&&
\times \, T^{-1} \big\{ \left[ M (\brm{x})  M (\brm{x}^\prime) \right]_C  - \left[ M (\brm{x})  \right]_C \left[ M (\brm{x}^\prime) \right]_C \big\}
\nonumber \\
&& = \, T^{-1} \big\{ \left[ M_{\text{tot}}^2 \right]_C - \left[ M_{\text{tot}} \right]_C^2 \big\} \, .
\end{eqnarray}
In the following we will refer to $\chi^{(1)}$ and $\chi^{(2)}$ in a brief fashion as susceptibilities.

\subsection{Review of the phase diagram}
\label{phase_diag_rev}
\begin{figure}
\begin{center}
\includegraphics[width=8cm]{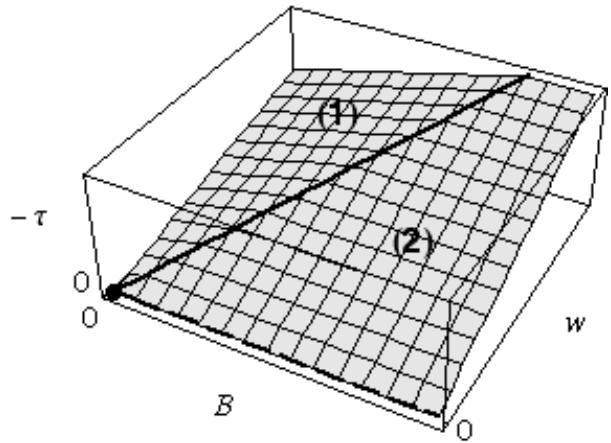}
\end{center}
\caption[]{\label{phaseDiag}Schematic phase diagram of the RJN. The critical surface (1) separates the insulating and the superconducting phase. The critical surface (2) lies between the insulating and the SG phase. Both surfaces merge in a line of bicritical points (solid) where the insulating, the superconducting, and the SG phase meet. For $w=0$ there is a critical line (dashed) separating the insulating and the SG phase. At the origin there is a critical point (dot) between the three phases.}
\end{figure}

In favor of a self contained presentation, we now briefly review the phase diagram of the RJN~\cite{john_lubensky_85_86}, see Fig~\ref{phaseDiag}. 
In mean-field theory, the phase diagram can be mapped out by determining those combinations of the parameters $p$, $T$, and $B$ for which the Gaussian part of $\mathcal{H}$ develops the eigenvalue zero. For convenience, we write the space coordinates $\brm{x} = (x,y, \brm{x}_\perp)$, where $\brm{x}_\perp$ lies in the $(d-2)$-dimensional subspace perpendicular to the $x$-$y$-plane. In the following we use the Landau gauge, i.e., we set $\brm{A} (\brm{x}) = (0, x, \brm{0}_\perp)$. The eigenvalues of the Gaussian part (the Landau levels) can be determined by standard textbook methods. One finds
\begin{eqnarray}
\label{landau}
E \big(  \brm{q}_\perp , \vec{\lambda}, m \big) = \tau + \brm{q}_\perp^2 + w \vec{\lambda}^2 + (2m+1) \big| \omega  \big(  \vec{\lambda} \big) \big| \, .
\end{eqnarray}
The momentum $\brm{q}_\perp$ is the Fourier transform of $\brm{x}_\perp$. $m$ labels the Landau levels and takes on the values $m = 0, 1, 2, \ldots$. $\omega (\vec{\lambda})$ is a cyclotron frequency given by
\begin{eqnarray}
\label{frequency}
\omega \big(  \vec{\lambda} \big) = \vec{B} \cdot \vec{\lambda} \, .
\end{eqnarray}
In the following we will assume that the external magnetic field is replica symmetric and write $B = B^{(\alpha )}$ for every $\alpha$.

Obviously, the lowest eigenvalue is associated with $\brm{q}_\perp = \brm{0}_\perp$. Now suppose that $w>0$ and $B>0$. By inspection one finds two modes associated with vanishing eigenvalues: (i) One with $\vec{\lambda}_1 = (1, 0, \ldots ,0)$ and $m=0$. It is soft, i.e., its eigenvalue vanishes, for $\tau + w + B =0$. (ii) One with $\vec{\lambda}_2 = (1, -1, \ldots ,0)$ and arbitrary $m$ that is soft for $\tau + 2w =0$ (note, that $\tau + w + B =\tau + 2w$ for $B =w$). $\vec{\lambda}_1$ and $\vec{\lambda}_2$ should be understood as representative for a set of equivalent (i.e., identical up to a permutation of the components) replica vectors. We identify two critical surfaces. The surface specified by $\tau = - (w+B)$ and  $B < w$ separates the insulating and the superconducting phase. The order parameter for this transition is $\langle \psi_{\vec{\lambda}_1}(\brm{x}) \rangle \stackrel{n \to 0}{\longrightarrow}  [ \langle e^{i \theta_{\brm{x}} } \rangle_T ]_C$. The surface corresponding to $\tau = -2w$ and $B>w$ separates the insulating and the SG phase. Here, the order parameter is of the Edwards-Anderson type, $\langle \psi_{\vec{\lambda}_2}(\brm{x}) \rangle \stackrel{n \to 0}{\longrightarrow} [ \langle e^{i \theta_{\brm{x}}} \rangle_T \langle e^{-i \theta_{\brm{x}}} \rangle_T ]_C$. The two critical surfaces merge at a line of bicritical points, given by $\tau = -2w = -2B$, where the insulating, the superconducting, and the SG phase meet. 

Now we come to the transitions I and II that are the main concern of this paper. Admitting $w=0$, one finds a critical line specified by $\tau = w =0$. Crossing over this line by tuning $\tau >0$ ($p<p_c$) to  $\tau <0$ ($p>p_c$) one gets from the insulating to the SG phase. This is our transition II. The order parameter here is $\langle \psi_{\vec{\lambda}}(\brm{x}) \rangle$ with all $\vec{\lambda} \neq \vec{0}$ satisfying $\sum_{\alpha =1}^n \lambda^{(\alpha )} = 0$. Finally, there is the critical point $\tau = w = B =0$ that represents the terminus of the two critical lines. Tuning $\tau >0$ to $\tau <0$ about this point one crosses over from the insulating to the superconducting phase. This is our transition I. Its order parameter is $\langle \psi_{\vec{\lambda}}(\brm{x}) \rangle$ with arbitrary $\vec{\lambda} \neq \vec{0}$. 

\section{Renormalization group analysis}
\label{renormalizationGroupAnalysis}

\subsection{Diagrammatic elements}
\label{diagrammaticElements}
To set up a diagrammatic perturbation expansion we need to identify the elements contributing to our Feynman diagrams. Evidently, there is the vertex $-g/T$. The Gaussian propagator for the present problem is not straightforward to determine, at least not in a non-approximate and closed form. That is due to the presence of the gauge field. A possible approach is to  expand the propagator in terms of the eigenfunctions belonging to the Landau eigenvalues~(\ref{landau}). This route was taken by Lawrie~\cite{lawrie_97} in studying the related problem of the LGW superconductor. By summing over all Landau levels Lawrie brought the LGW propagator into an elegant form that made multi-loop calculations tractable. In Appendix~\ref{app:propagator} we present a simpler approach that allows us to derive the propagator directly without resorting to a expansion in terms of Landau levels. Yet, our approach reproduces the closed form found by Lawrie. We obtain
\begin{subequations}
\label{boldProp}
\begin{eqnarray}
&&G^{\text{bold}} \big(  \brm{x}, \brm{x}^\prime , \vec{\lambda}, \big) = T \exp \left[  i \frac{\omega \big(  \vec{\lambda} \big)}{2}    \left( x + x^\prime \right) \left(  y - y^\prime \right)  \right]
\nonumber \\
&& \times \, \int_{\brm{k}} \widetilde{G}  \big(  \brm{k}, \vec{\lambda} \big) \exp \left[  i  \brm{k} \cdot \left(   \brm{x} - \brm{x}^\prime \right) \right]  \left(  1 - \delta_{\vec{\lambda}, \vec{0}} \right)
\end{eqnarray}
as the principal propagator for the RJN. Here, $ \int_{\brm{k}}$ is the usual shorthand notation for $1/(2\pi )^d \int d^d k$. $ \widetilde{G}  (  \brm{k}, \vec{\lambda} )$ is given by
\begin{eqnarray}
\label{resGtilde}
&& \widetilde{G} \big(  \brm{k} , \vec{\lambda} \big) =  \int_0^\infty \frac{ds}{\cosh \left(   \omega  \big(  \vec{\lambda} \big) s \right)}
\nonumber \\
&& \times \,  \exp \left[  -s   \left( \tau  ( \vec{\lambda} ) +  \brm{k}_\perp^2 \right) -  \frac{\tanh \left(   \omega  \big(  \vec{\lambda} \big) s \right)}{ \omega  \big(  \vec{\lambda} \big)} \left(   p^2 + q^2 \right)  \right]  \, .
\nonumber \\
\end{eqnarray}
\end{subequations}
where $p$ and $q$ are conjugate to $x$ and $y$, respectively. $\tau (\vec{\lambda})$ is a shorthand notation for $\tau + w \vec{\lambda}^2$. The factor $(  1 - \delta_{\vec{\lambda}, \vec{0}} )$ implements the constraint $ \vec{\lambda} \neq  \vec{0}$. 

We annotate that one could discuss the RJN phase diagram by analyzing the infrared behavior of the propagator~(\ref{boldProp}) instead of the minima of the Landau levels~(\ref{landau}).  Basically, one just has to consider the limit $s \to \infty$ and to determine the parameter combinations for which the propagator becomes long range. Of course, one finds the phase diagram discussed in Sec.~\ref{phase_diag_rev}

We observe that the principal propagator decomposes into two parts,
\begin{eqnarray}
\label{deco}
G^{\text{bold}} \big(  \brm{x}, \brm{x}^\prime , \vec{\lambda} \big) = G^{\text{cond}} \big(  \brm{x}, \brm{x}^\prime , \vec{\lambda} \big) - G^{\text{ins}} \big(  \brm{x} - \brm{x}^\prime \big) \, .
\nonumber \\
\end{eqnarray}
One of them 
\begin{eqnarray}
\label{condProp}
&&G^{\text{cond}} \big(  \brm{x}, \brm{x}^\prime , \vec{\lambda} \big) = T \exp \left[  i \frac{\omega \big(  \vec{\lambda} \big)}{2}    \left( x + x^\prime \right) \left(  y - y^\prime \right)  \right]
\nonumber \\
&& \times \, \int_{\brm{k}} \widetilde{G}  \big(  \brm{k}, \vec{\lambda} \big) \exp \left[  i  \brm{k} \cdot \left(   \brm{x} - \brm{x}^\prime \right) \right]  \, ,
\end{eqnarray}
carries $\vec{\lambda}$. The other one, 
\begin{eqnarray}
\label{insProp}
G^{\text{ins}} \big(  \brm{x} - \brm{x}^\prime \big) =  T \int_{\brm{k}} \,  \frac{ \delta_{\vec{\lambda}, \vec{0}}}{\tau + \brm{k}^2} \, \exp \left[  i  \brm{k} \cdot \left(   \brm{x} - \brm{x}^\prime \right) \right]  \, ,
\end{eqnarray}
does not. The notation we use here reflects the close analogy to the RRN, where $\vec{\lambda}$ plays essentially the role of a current. For the RRN the decomposition of the principal propagator culminates in a real world interpretation~\cite{stenull_janssen_oerding_99,stenull_2000,janssen_stenull_oerding_99,janssen_stenull_99,stenull_janssen_epl_2000,stenull_janssen_2001,stenull_janssen_oerding_2001,janssen_stenull_prerapid_2001,stenull_janssen_jsp_2001,stenull_janssen_pre_2001_nonlinear,janssen_stenull_pre_2001_vulcanization,stenull_janssen_epl_2001,janssen_stenull_pre_2001_continuum,stenull_janssen_pre_2002} in that the Feynman diagrams are viewed as conducting networks composed of insulating and conducting propagators. The insulating propagator ~(\ref{insProp}) is identical in form to its counterpart for the RRN. The conducting propagator ~(\ref{condProp}) reduces formally to its analog for the RRN for vanishing cyclotron frequency,
\begin{eqnarray}
\label{reducedProp}
&&G^{\text{bold}} \big(  \brm{x}, \brm{x}^\prime , \vec{\lambda} \big) \stackrel{\omega \to 0}{\longrightarrow}  T \int_{\brm{k}}  \frac{\exp \left[  i  \brm{k} \cdot \left(   \brm{x} - \brm{x}^\prime \right) \right]}{\tau + w  \vec{\lambda}^2 + \brm{k}^2} \nonumber \\
&&= \, G \big(  \brm{x} - \brm{x}^\prime , \vec{\lambda} \big) \, .
\end{eqnarray}
Comparing $G (  \brm{x} - \brm{x}^\prime , \vec{\lambda} )$ to the full conducting propagator $G^{\text{cond}} (  \brm{x}, \brm{x}^\prime , \vec{\lambda} )$ it is apparent that the perturbation theory simplifies tremendously for vanishing $\omega (\vec{\lambda})$. This simplification will allow us to study the susceptibilities at transition I and II with reasonable effort.

\subsection{Order parameter correlation functions}
\label{opCorr}
Here we will discuss the renormalization and the scaling behavior of the order parameter correlation functions
\begin{eqnarray}
G_N \big(  \big\{  \brm{x},  \vec{\lambda} \big\}; \tau ,w , B, g \big) = \left\langle  \psi_{\vec{\lambda}_1}(\brm{x}_1 ) , \ldots , \psi_{\vec{\lambda}_N }(\brm{x}_N )\right\rangle \, ,
\end{eqnarray}
where we drop the redundant scaling variable $T$ for notational simplicity. Though these correlation functions are not the main concern of this paper, they deserve some attention. First, they are interesting in their own right. Second, their discussion will provide some background for the subsequent analysis of the susceptibilities.  Most of the techniques we are going to use, such as dimensional regularization and minimal subtraction, belong to the standard repertoire of renormalized field theory, cf.\ Rev.~\cite{amit_zinn-justin}. 

To remove ultraviolet (UV) divergences from the order parameter correlation functions finite, we use the renormalization scheme 
\begin{subequations}
\label{renScheme}
\begin{eqnarray}
\psi \to \mathring{\psi} &=& Z^{1/2} \psi \, ,
\\
\tau \to \mathring{\tau} &=& Z^{-1} Z_{\tau} \tau  \, ,
\\
w \to \mathring{w} &=& Z^{-1} Z_w w \, , 
\\
g \to \mathring{g} &=& Z^{-3/2} Z_u^{1/2} G_\varepsilon^{-1/2} u^{1/2} T^{-1/2} \mu^{\varepsilon/2} \, ,
\\
B \to \mathring{B} &=& Z^{-1/2} Z_B B
\end{eqnarray}
\end{subequations}
where the $\mathring{}$ indicates unrenormalized quantities. $\mu$ is the usual inverse length scale. $\varepsilon = 6-d$ specifies the deviation from the upper critical dimension 6. The factor $G_\varepsilon = (4\pi )^{-d/2}\Gamma (1 + \varepsilon /2)$ is introduced for later convenience.

At first, we consider the role of the magnetic field. For the closely related problem of the LGW superconductor it was demonstrated by Lawrie~\cite{lawrie_97} in a two-loop calculation that the $B$-field does not require renormalization, i.e., that $\mathring{B} =B$ up to second order in $\varepsilon$-expansion, and that the remaining $Z$-factors are independent of $B$. One can show, however, that the validity of these findings is not limited to second order; they are valid to arbitrary order in $\varepsilon$-expansion. We will address these points in some detail in a forthcoming publication~\cite{janssen_stenull_future} on the LGW superconductor. The quintessential points can be sketched as follows: the non-renormalization of $B$ is a consequence of the gauge invariance. Exploiting this invariance, one can show that the current density renormalizes trivially. Because the vector potential  is conjugate to the current density, it does not  require an independent renormalization factor also. In turn, $B$ renormalizes trivially, i.e., $Z^{-1/2} Z_B =1$ to arbitrary order in $\varepsilon$-expansion.  The $B$-independence of the $Z$-factors follows from the fact that $B$ is not dimensionless at the upper critical dimension.

For the present problem the reasoning of Ref.~\cite{janssen_stenull_future} has to be modified somewhat. That is because the magnetic field appears in case of the RJN always in the combination $w^{-1/2} B \sim \mu$ as opposed to the pure $B \sim \mu^2$ in case of the LGW superconductor. As a consequence, the mass $\tau$ requires in the present problem an additive renormalization proportional to $B^2$. This subtlety has no consequence for our main results and we will ignore it in the following.

The most economic way to determine $Z$,  $Z_\tau$, $Z_w$, and $Z_g$ is to exploit the close relation of the RJN to the RRN. As indicated earlier, the corresponding two diagrammatic expansions coincide for $B=0$ in the replica limit. Furthermore, the RRN reduces for $w=0$ to purely geometrical percolation. Hence,  $Z$,  $Z_\tau$, and $Z_g$ are nothing but the usual percolation $Z$-factors known to third order in $\varepsilon$-expansion~\cite{alcantara_80}.  $Z_w$  may be gleaned to second order  in $\varepsilon$  from our work on the RRN~\cite{stenull_janssen_oerding_99,stenull_2000}.

Having determined the $Z$ factors, we are now in the position to infer the scaling behavior of the order parameter correlation functions from their renormalization group equation (RGE). This RGE is a manifestation of the fact that the unrenormalized theory has to be independent of the arbitrary length scale $\mu^{-1}$ introduced by renormalization. Hence, the unrenormalized correlation functions satisfy the identity
\begin{eqnarray}
\label{independence}
\mu \frac{\partial}{\partial \mu} \mathring{G}_N \left( \left\{ {\rm{\bf x}} ,\vec{\lambda} \right\} ;   \mathring{\tau}, \mathring{w},  \mathring{B}, \mathring{g} \right) = 0 \, . \end{eqnarray}
Equation~(\ref{independence}) translates via the Wilson functions
\begin{subequations}
\label{wilson}
\begin{eqnarray}
\label{wilsonGamma}
\gamma_{\ldots } \left( u \right) &=& \mu \frac{\partial }{\partial \mu} \ln Z_{\ldots } \bigg|_0 \, ,
\\
\label{betau}
\beta \left( u \right) &=& \mu \frac{\partial u}{\partial \mu} \bigg|_0 = u \left( 3 \gamma - \gamma_u - \varepsilon \right) \, ,
\\
\kappa \left( u \right) &=& \mu \frac{\partial \ln \tau}{\partial \mu} \bigg|_0 = \gamma - \gamma_\tau \, ,
\\
\label{wilsonZeta}
\zeta \left( u \right) &=& \mu \frac{\partial \ln w}{\partial \mu} \bigg|_0 = \gamma - \gamma_w \, ,
\end{eqnarray}
\end{subequations}
(the $|_0$ indicates that bare quantities are kept fix while taking the derivatives) into the RGE
\begin{eqnarray}
&&\left[ \mu \frac{\partial }{\partial \mu} + \beta \frac{\partial }{\partial u} + \tau \kappa \frac{\partial }{\partial \tau} + w \zeta 
\frac{\partial }{\partial w} + \frac{N}{2} \gamma \right] 
\nonumber \\
&&\times \, 
G_N \left( \left\{ {\rm{\bf x}} ,\vec{\lambda} \right\} ; \tau, w, B, u, \mu \right) = 0 \, .
\end{eqnarray}

The RGE can be solved in terms of a single flow parameter $\ell$ by using the characteristics
\begin{subequations}
\begin{eqnarray}
\ell \frac{\partial \bar{\mu}}{\partial \ell} &=& \bar{\mu} \, , \quad \bar{\mu}(1)=\mu \ ,
 \\
\label{charBeta}
\ell \frac{\partial \bar{u}}{\partial \ell} &=& \beta \left( \bar{u}(\ell ) \right) \, , \quad \bar{u}(1)=u \, ,
 \\
\ell \frac{\partial}{\partial \ell} \ln \bar{\tau} &=& \kappa \left( \bar{u}(\ell ) \right) \, , \quad \bar{\tau}(1)=\tau \, ,
 \\
\ell \frac{\partial}{\partial \ell} \ln \bar{w} &=& \zeta \left( \bar{u}(\ell ) \right) \, , \quad \bar{w}(1)=w \, ,
 \\
\ell \frac{\partial}{\partial \ell} \ln \bar{Z} &=& \gamma \left( \bar{u}(\ell ) \right) \, , \quad \bar{Z}(1)=1 \, .
\end{eqnarray}
\end{subequations}
These characteristics describe how the parameters transform if we change the momentum scale $\mu $ according to $\mu \to \bar{\mu}(\ell )=\ell\mu $. Being interested in the infrared (IR) behavior of the theory, we study the limit $\ell \to 0$. According to Eq.~(\ref{charBeta}) we expect that in this IR limit the coupling constant $\bar{u}(\ell )$ flows to a stable fixed point $u^\ast$ satisfying $\beta (u^\ast )=0$. The IR stable fixed point solution to the RGE is readily found. In conjunction with dimensional analysis (to account for naive dimensions) it gives 
\begin{eqnarray}
\label{scalingOPcorr}
&&G_N \left( \left\{ {\rm{\bf x}} ,\vec{\lambda} \right\} ;  \tau , w, B, u, \mu \right) = 
\ell^{(d-2+\eta)N/2} 
\nonumber \\
&& \times \,
G_N \left( \left\{ \ell {\rm{\bf x}} , \vec{\lambda} \right\} ;  \ell^{-1/\nu}\tau , \ell^{-\phi /\nu} w, \ell^{-2} B, u^\ast, \mu \right)
\nonumber \\
\end{eqnarray}
with the critical exponents for percolation $\eta = \gamma (u^\ast )$ and $\nu = [2-\kappa (u^\ast)]^{-1}$ known to third order in $\varepsilon$~\cite{alcantara_80}. $\phi = \nu [2- \zeta (u^\ast )]$ is the percolation resistance exponent known to second order in $\varepsilon$~\cite{lubensky_wang_85,stenull_janssen_oerding_99,stenull_2000}. We have not yet exploited the freedom to choose $\ell$. By setting for example $\ell = |\tau |^\nu$ we find that the order parameter correlation functions scale like
\begin{eqnarray}
\label{scalingOPcorrFinal}
&&G_N \left( \left\{ {\rm{\bf x}} ,\vec{\lambda} \right\} ;  \tau , w, B, u, \mu \right) = 
\xi^{-(d-2+\eta)N/2} 
\nonumber \\
&& \times \,
\Omega_N \left( \left\{ \xi^{-1} {\rm{\bf x}} , \xi^{\phi / 2 \nu} w^{1/2} \vec{\lambda} \right\} ; \xi^{2 - \phi / 2 \nu} w^{- 1/2} B \right) \, ,
\nonumber \\
\end{eqnarray}
where $\xi$ is the correlation length and $\Omega_N$ is a scaling function.  

\subsection{Susceptibilities}
In this section we discuss the renormalization of the susceptibilities as well as their scaling behavior for transition I and II. We start by analyzing the Feynman diagrams contributing to the susceptibilities. First, we consider the one-loop order in some detail. More important than yielding concrete results, this provides us with some intuition about the general structure of the diagrams. Then, this general structure is analyzed for diagrams with an arbitrary number of loops. We demonstrate how to renormalize the susceptibilities properly. Finally, we derive their scaling behavior. 

\subsubsection{Diagrammatics: one-loop calculation}
\label{diagramatics:One-loopCalculation}
Our starting point here is the definition of the current density correlation function $C_{i,j}^{(\alpha ,\beta)} (\brm{x} - \brm{x}^\prime)$ in Eq.~(\ref{currCorr}). At first, we express this correlation function directly in terms of the order parameter field. Recall that $C_{i,j}^{(\alpha ,\beta)} (\brm{x} - \brm{x}^\prime)$ is defined originally in terms of the current density 
\begin{eqnarray}
\label{currdensA0}
&&J_i^{(\alpha )} (\brm{x} ) = \sum_{\vec{\lambda} \neq \vec{0}} \frac{1}{2i} \,  \lambda^{(\alpha )} 
\nonumber \\
&& \times \, \big[  \psi_{-\vec{\lambda}} \left( {\rm{\bf x}} \right) \partial_i  \psi_{\vec{\lambda}} \left( {\rm{\bf x}} \right) -  \psi_{\vec{\lambda}} \left( {\rm{\bf x}} \right) \partial_i  \psi_{-\vec{\lambda}} \left( {\rm{\bf x}} \right)\big] \, ,
\end{eqnarray}
where we have set $\vec{\brm{A}} = \vec{\brm{0}}$. Being a composite field, the current density is inconvenient to handle in actual calculations. Hence, we substitute (\ref{currdensA0}) into the current density correlation function. Then $C_{i,j}^{(\alpha ,\beta)} (\brm{x} - \brm{x}^\prime)$ is composed directly of correlation functions of the order parameter field and its derivatives. Next, we expand the weight $\exp (- T^{-1}\mathcal{H})$ in powers of the coupling constant $g/T$. At zeroth order in $g/T$ we obtain
\begin{eqnarray}
\label{progress}
&& C_{i,j}^{(\alpha ,\beta)} (\brm{x} - \brm{x}^\prime) =  \sum_{\vec{\lambda}} \frac{1}{2i} \,  \lambda^{(\alpha )} \lambda^{(\beta )} 
\nonumber \\
&& \times \, 
\Big\{  T^{-1} \big[G^{\text{bold}} \big(  \brm{x}^\prime, \brm{x} , \vec{\lambda} \big) \, \partial_i \partial_j^\prime G^{\text{bold}} \big(  \brm{x}, \brm{x}^\prime , \vec{\lambda} \big)
\nonumber \\
&& - \, \partial_i  G^{\text{bold}} \big(  \brm{x}, \brm{x}^\prime , \vec{\lambda} \big)  \, \partial_j^\prime  G^{\text{bold}} \big(  \brm{x}^\prime, \brm{x} , \vec{\lambda} \big) \big]
\nonumber \\
&&
-  \, \delta \big(  \brm{x} - \brm{x}^\prime \big)  \delta_{i,j} \, G^{\text{bold}} \big(  \brm{0}, \vec{\lambda} \big)
\Big\} \, .
\end{eqnarray}
Here it is understood that the bold propagator is evaluated at $\omega \big( \vec{\lambda} \big) = 0$ since we focus on transition I and II. Now we decompose the bold propagator into its conducting and its insulating part. We recall that the insulating propagator contains a factor $\delta_{\vec{\lambda}, \vec{0}}$. Due to the $\lambda^{(\alpha )} \lambda^{(\beta )}$ in (\ref{progress}) the all terms containing insulating propagators drop out. We obtain upon sending $\brm{x}^\prime \to \brm{0}$
\begin{eqnarray}
\label{progress2}
&& C_{i,j}^{(\alpha ,\beta)} (\brm{x} ) =  \sum_{\vec{\lambda}} \frac{1}{2i} \,  \lambda^{(\alpha )} \lambda^{(\beta )} 
 \Big\{  T^{-1} \big[ G \big(  \brm{x}, \vec{\lambda} \big) \, \partial_i \partial_j G \big(  \brm{x} , \vec{\lambda} \big)
\nonumber \\
&& - \, \partial_i G \big(  \brm{x}, \vec{\lambda} \big) \, \partial_j G \big(  \brm{x} , \vec{\lambda} \big) \big]
 -   \delta \big(  \brm{x}  \big)  \delta_{i,j} \, G \big(  \brm{0}, \vec{\lambda} \big)
\Big\} \, .
\end{eqnarray}

Next we switch to momentum space. Applying the usual Fourier transformation to Eq.~(\ref{progress2}) yields
\begin{eqnarray}
\label{progress3}
&& \widetilde{C}_{i,j}^{(\alpha ,\beta)} (\brm{k} ) = T  \sum_{\vec{\lambda}}   \lambda^{(\alpha )} \lambda^{(\beta )}  \int_{\brm{q}}
\nonumber \\
&& \times \,  \Bigg\{  \frac{2 \, q_i q_j}{\big[ \tau \big( \vec{\lambda} \big) + (\brm{q} + \brm{k}/2)^2 \big] \big[ \tau \big( \vec{\lambda} \big) + (\brm{q} - \brm{k}/2)^2 \big]} 
\nonumber \\
&& - \, \frac{\delta_{i,j}}{\tau \big( \vec{\lambda} \big) + \brm{q}^2}
\Bigg\} \, .
\end{eqnarray}
Since we are ultimately interested in the susceptibilities, we should look at $\widetilde{C}^{(\alpha ,\beta)} (\brm{k} )$ rather than $\widetilde{C}_{i,j}^{(\alpha ,\beta)} (\brm{k} )$. Via taking the trace on both sides of Eq.~(\ref{corroFourier}) we find that
\begin{eqnarray}
\label{progress4}
&& \widetilde{C}^{(\alpha ,\beta)} (\brm{k} ) =  
\frac{2 \, T}{(d-1) \, \brm{k}^2} \sum_{\vec{\lambda}}   \lambda^{(\alpha )} \lambda^{(\beta )}  \int_{\brm{q}} \brm{q}^2
\nonumber \\
&& \times \,  \Bigg\{  \frac{1}{\big[ \tau \big( \vec{\lambda} \big) + (\brm{q} + \brm{k}/2)^2 \big] \big[ \tau \big( \vec{\lambda} \big) + (\brm{q} - \brm{k}/2)^2 \big]} 
\nonumber \\
&& - \, \frac{1}{\big[ \tau \big( \vec{\lambda} \big) + \brm{q}^2 \big]^2}
\Bigg\} \, .
\end{eqnarray}
Here we have used that
\begin{eqnarray}
\int_{\brm{q}} \frac{d}{\tau \big( \vec{\lambda} \big) + \brm{q}^2 } = \int_{\brm{q}} \frac{1}{\big[ \tau \big( \vec{\lambda} \big) + \brm{q}^2 \big]^2} 
\end{eqnarray}
in dimensional regularization. By virtue of the relation~(\ref{relChiC}) we obtain upon expansion in the external momentum $\brm{k}$
\begin{eqnarray}
\label{progress5}
\chi^{(\alpha ,\beta)}  =  -
\frac{T}{6} \sum_{\vec{\lambda}}  \int_{\brm{q}}  \frac{ \lambda^{(\alpha )} \lambda^{(\beta )}}{\big[ \tau \big( \vec{\lambda} \big) + \brm{q}^2 \big]^2} \, .
\end{eqnarray}
This is the replica susceptibility at one-loop order.

We proceed with simplifying the summation over $\vec{\lambda}$. In Schwinger representation, this summation is of the form
\begin{eqnarray}
\sum_{\vec{\lambda}} \lambda^{(\alpha )} \lambda^{(\beta )}  \exp \left[  -s \tau \big( \vec{\lambda} \big) \right] \, .
\end{eqnarray}
At this point, we find it convenient to introduce the shorthand notation
\begin{eqnarray}
\langle \ldots \rangle_\lambda = \sum_{\vec{\lambda}} \ldots \exp \left[  -s \tau \big( \vec{\lambda} \big) \right]  \, .
\end{eqnarray}
Revisiting Eq.~(\ref{propChi}), we deduce that 
\begin{eqnarray}
\langle  \lambda^{(\alpha )} \lambda^{(\beta )} \rangle_\lambda =  a \, \delta_{\alpha ,\beta} + b \, ,
\end{eqnarray}
where $a$ and $b$ are coefficients that need to be determined. By analyzing the cases $\alpha = \beta$ and $\alpha \neq \beta$ we find
\begin{eqnarray}
a = \frac{1}{n(n-1)} \bigg[  n \left\langle  \vec{\lambda}^2 \right\rangle_\lambda - \bigg\langle  \bigg( \sum_{\alpha =1}^n \lambda^{(\alpha )}  \bigg)^2  \bigg\rangle_\lambda \bigg]  \, ,
\\
b = - \frac{1}{n(n-1)} \bigg[  \left\langle  \vec{\lambda}^2 \right\rangle_\lambda - \bigg\langle \bigg( \sum_{\alpha =1}^n \lambda^{(\alpha )}  \bigg)^2  \bigg\rangle_\lambda \bigg]  \, .
\end{eqnarray}
To evaluate $a$ and $b$ further we now look at transition I and II separately. At the transition I we have $B=0$. For $B=0$, the system is, like the RRN, rotationally invariant in replica space and hence $\langle  \left( \sum_{\alpha =1}^n \lambda^{(\alpha )}  \right)^2  \rangle_\lambda = \langle  \vec{\lambda}^2 \rangle_\lambda$. Consequently, we  have $a = \frac{1}{n} \langle  \vec{\lambda}^2 \rangle_\lambda$ and $b=0$. This leads for the susceptibilities at transition I to
\begin{eqnarray}
\label{progress6}
\chi^{(1)}  =  -
\frac{T}{6 n} \sum_{\vec{\lambda}}  \int_{\brm{q}}  \frac{\vec{\lambda}^2}{\big[ \tau \big( \vec{\lambda} \big) + \brm{q}^2 \big]^2} \, , \  \
\chi^{(2)}  = 0 \, .
\end{eqnarray}
At transition II we have $\sum_{\alpha =1}^n \lambda^{(\alpha )} = 0$ and thus $a = O (n^0)$ and $b = \frac{1}{n} \langle  \vec{\lambda}^2 \rangle_\lambda$. For the susceptibilities at transition II this leads to
\begin{eqnarray}
\label{progress7}
\chi^{(1)}  = 0 \, , \ \
\chi^{(2)}  = - \frac{T}{6 n} \sum_{\vec{\lambda}}  \int_{\brm{q}}  \frac{\vec{\lambda}^2 \delta_{\sum_{\alpha =1}^n \lambda^{(\alpha )} , 0}}{\big[ \tau \big( \vec{\lambda} \big) + \brm{q}^2 \big]^2}  \, .
\end{eqnarray}

Now we are in the position to carry out the remaining summations over $\vec{\lambda}$ along with the momentum integration. We outline the remaining steps at the instance of $\chi^{(2)}$ at transition II. Implementing the constraint $\sum_{\alpha =1}^n \lambda^{(\alpha )} = 0$ via the integral $\frac{1}{2\pi} \int_{-\pi}^\pi d \theta \, e^{i \theta \sum_{\alpha =1}^n \lambda^{(\alpha )}}$ we have
\begin{eqnarray}
\label{progress8}
\chi^{(2)}  =  -
\frac{T}{6 n} \sum_{\vec{\lambda}}  \frac{1}{2\pi} \int_{-\pi}^\pi d \theta \, e^{i \theta \sum_{\alpha =1}^n \lambda^{(\alpha )}} \int_{\brm{q}}  \frac{\vec{\lambda}^2}{\big[ \tau \big( \vec{\lambda} \big) + \brm{q}^2 \big]^2} \, .
\nonumber \\
\end{eqnarray}
We find it convenient to use the Schwinger representation and rewrite $\chi^{(2)}$ as
\begin{eqnarray}
\label{progress9}
&& \chi^{(2)}  =  -
\frac{T}{6 n} \int_0^\infty ds \, s  \,  \frac{1}{2\pi} \int_{-\pi}^\pi d \theta \,  \int_{\brm{q}}  \frac{\partial}{w \, \partial z} \sum_{\vec{\lambda}}   
\nonumber \\
&& \times \, \exp \left[ x w \vec{\lambda}^2 - s \left(  \tau + w \vec{\lambda}^2 + \brm{q}^2 \right) +  i \theta \sum_{\alpha =1}^n \lambda^{(\alpha )} \right] \Bigg|_{z=0} \, .
\nonumber \\
\end{eqnarray}
At this point it is useful to exploit Poisson's summation formula and approximate the summation $\sum_{\vec{\lambda}}$ by the integration $\int_{-\infty}^\infty d^n \lambda$. This integration is Gaussian and it can be evaluated straightforwardly by completing squares in the exponential. Next, we carry out the particularly simple momentum integration. Since we eventually have to take the replica limit $n\to0$, we then expand  $\chi^{(2)}$ in powers of $n$. Upon taking the derivative with respect to $z$ and carrying out the integration over $\theta$ we get in the replica limit
\begin{eqnarray}
\label{progress10}
\chi^{(2)}  =  - \frac{T}{12} \int_0^\infty ds \,   \frac{e^{-s\tau}}{(4\pi s)^{d/2}}  \bigg\{ - \frac{1}{w} + \frac{\pi^2}{6sw^2} \bigg\}   \, .
\end{eqnarray}
The remaining integration over the Schwinger parameter $s$ represents no difficulty. For the renormalization group treatment that we have in mind we will only need the UV divergent parts of $\chi$. These are extracted in form of $\varepsilon$-poles by expanding $\chi$ in powers of $\varepsilon$. This provides us finally with the following divergent parts of $\chi^{(2)}$ for transition II:
\begin{eqnarray}
\label{progress11}
\chi^{(2)}_{\text{div}}  =    - \frac{T \, \tau^{2}}{w} \, \frac{ G_\varepsilon}{12 \, \varepsilon}  \bigg\{ 1 +  \frac{\pi^2}{18} \frac{\tau}{w} \bigg\}  \, .
\end{eqnarray}
Note that the 1-loop result for $\chi^{(2)}_{\text{div}}$ given in Rev.~\cite{JLW_88} is erroneous. It incorrectly features $w_2$.

$\chi^{(1)}$ for transition I can be calculated in an analogous manner. We merely need to replace the integration that enforces the constraint $\sum_{\alpha =1}^n \lambda^{(\alpha )} = 0$ by unity. This leads to the result
\begin{eqnarray}
\label{progress12}
\chi^{(1)}_{\text{div}}  =  - \frac{T \, \tau^{2}}{w} \, \frac{ G_\varepsilon}{12 \, \varepsilon}  \, .
\end{eqnarray}
At this point a comment on the $\tau$ dependence of the susceptibilities is in order.  From a technical point of view, dimensional regularization is the most convenient way of dealing with UV divergences. However, dimensional regularization has some unphysical features which are intimately related to its simplicity. In the less economic but more physical cutoff regularization one treats UV singularities by introducing a cutoff $\Lambda$ and by replacing the full integration $\int_{\brm{q}}$ by  $\int_{ |\brm{q} | \leq \Lambda}$. This procedure leads typically to terms proportional to $\ln \Lambda$ and terms varying as some power of $\Lambda$. The logarithmic divergences for $\Lambda \to \infty$ have their analog in dimensional regularization in form of the $\varepsilon$ poles. The terms algebraic in $\Lambda$, however, are unaccounted for in dimensional regularization. In case of the susceptibilities this neglect conceals essential physics. Hence, the terms algebraic in $\Lambda$ must be incorporated. We get
\begin{eqnarray}
\label{LambdaIncorp1}
\chi^{(1)}_{\text{div}}  =  \frac{T}{w} \Big[ A_0 \Lambda^4 + A_1 \Lambda^2 \tau - \frac{G_\varepsilon}{12 \, \varepsilon} \, \tau^{2} \Big]
\end{eqnarray}
and 
\begin{eqnarray}
\label{LambdaIncorp2}
&& \chi^{(2)}_{\text{div}}  =  \frac{T}{w} \Big[ A_0 \Lambda^4 + A_1 \Lambda^2 \, \tau - \frac{G_\varepsilon}{12 \, \varepsilon} \, \tau^{2} \Big]
\nonumber \\
&& + \, \frac{T}{w^2} \Big[ B_0 \Lambda^6 + B_1 \Lambda^4 \, \tau + B_2 \Lambda^2 \, \tau^2 - \frac{\pi^2 G_\varepsilon}{12 \cdot 18 \, \varepsilon} \, \tau^{3} \Big] \, 
\nonumber \\
\end{eqnarray}
where the $A$'s and $B$'s are numerical coefficients.

\subsubsection{Diagrammatics: general structure}
\label{diagramatics:GeneralStructure}
Higher orders in the expansion of the the weight $\exp (- \mathcal{H})$ in powers of the coupling constant $g$ correspond to multi-loop diagrams. In Schwinger representation, the $\vec{\lambda}$-featuring part of such a diagram is of the form
\begin{eqnarray}
\label{start}
\sum_{\{  \vec{\kappa} \}}  \lambda^{(\alpha)}  \lambda^{\prime (\beta)}  \exp \left[ - \sum_p  s_p w  \vec{\lambda}^2_p \right]   \, .
\end{eqnarray}
Here, $\{  \vec{\kappa} \}$ stands for a complete set of independent loop currents. The sums over $p$ are taken over all conducting propagators. The currents $\vec{\lambda}_p$ running through the conducting propagators are linear functions of the loop currents, $\vec{\lambda}_p = \vec{\lambda}_p (\{  \vec{\kappa} \})$. The currents $\vec{\lambda}$ and $\vec{\lambda}^\prime$ are identical to particular $\vec{\lambda}_p$. In the one-loop example given in the preceding section, there is one loop current $\vec{\lambda}$ and the two $\vec{\lambda}_p$ as well as $\vec{\lambda}^\prime$ are  identical to $\vec{\lambda}$. 

\begin{figure}
\begin{center}
\includegraphics[width=8cm]{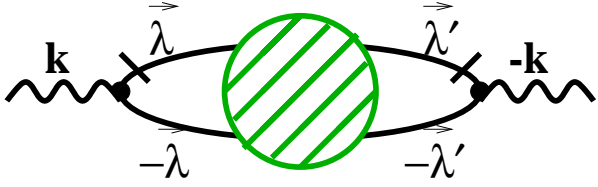}
\end{center}
\caption[]{\label{diagramStruct}General structure of the Feynman diagrams contributing to the susceptibilities. The hatched blob symbolizes an arbitrary number of closed loops composed of the vertex $-g$ and conducting and insulating propagators. To the left and the right of the blob we have conducting propagators. The wavy lines indicate insertion points for external momenta into these propagators. The two bars stand for a factor $\vec{\lambda} \cdot \vec{\lambda}^\prime$.}
\end{figure}

Of course, (\ref{start}) splits up into a replica diagonal and a replica independent part. Generalizing the arguments given in Sec.~\ref{diagramatics:One-loopCalculation} it is not difficult to deduce from (\ref{start}) that $\chi^{(2)} $ vanishes at transition I to arbitrary order in the loop expansion. The same goes for $\chi^{(1)}$ at transition II.  Moreover, one finds that the $\vec{\lambda}$-featuring part of non-vanishing diagrams is of the structure
\begin{eqnarray}
\label{struct1}
&&\frac{1}{n} \sum_{\{  \vec{\kappa} \}}  \vec{\lambda} \cdot \vec{\lambda}^\prime \exp \left[ - \sum_p \left( s_p w  \vec{\lambda}^2_p - i \theta_p \sum_{\alpha =1}^n \lambda^{(\alpha )}_p \right) \right] 
\nonumber \\
&& = \, \frac{1}{n} \frac{\partial}{w \, \partial z} \sum_{\{  \vec{\kappa} \}}   \exp \Bigg[  z w  \vec{\lambda} \cdot \vec{\lambda}^\prime   
\nonumber \\
&&
- \, \sum_p \Bigg( s_p w  \vec{\lambda}^2_p -  i \theta_p \sum_{\alpha =1}^n \lambda^{(\alpha )}_p \Bigg) \Bigg]  \Bigg|_{z=0}  \, .
\end{eqnarray}
In case of the transition I the $\theta_p$ are all zero. Figure~\ref{diagramStruct} depicts these non-vanishing diagrams.

Now to the evaluation of Eq.~(\ref{struct1}). Its right hand side factorizes into $n$ equivalent factors. Exploiting the fact that the $\vec{\lambda}_p$ are linear functions of the loop currents, we write
\begin{eqnarray}
\label{struct2}
&&\frac{1}{n} \frac{\partial}{w \, \partial z} \prod_{\alpha =1}^n \sum_{\{  \kappa^{(\alpha )} \}}   \exp \bigg[  -w \sum_{l, l^\prime}   \kappa^{(\alpha )}_l A_{l, l^\prime} (\{s\} , z) \,  \kappa^{(\alpha )}_{l^\prime}
\nonumber \\
&&
+ \,  i \sum_l b_l (\{ \theta \} )  \, \kappa^{(\alpha )}_l \bigg]  \bigg|_{z=0}  
\end{eqnarray}
for the right hand side of Eq.~(\ref{struct1}). The sums over $l$ and $l^\prime$ run over the complete set of independent conducting loops corresponding to $\{  \vec{\kappa} \}$. $A_{l, l^\prime} (\{s\} , z)$ and $b_l (\{ \theta \} )$ are linear functions of their variables. In case of the transition I the $b_l (\{ \theta \} )$ are all zero. Since we are interested in the limit $w \to 0$, we may apply Poisson's summation formula, i.e.,  replace the summations $\sum_{\{  \kappa^{(\alpha )} \}}$ by integrations $\int_{-\infty}^\infty d \kappa^{(\alpha )}$. The so obtained integrations are Gaussian and hence straightforward. They yield
\begin{eqnarray}
\label{struct3}
\frac{1}{n} \frac{\partial}{w \, \partial z}  \bigg[  \frac{1}{\sqrt{\det (w \underline{\underline{A}})}} \exp \left(  - \frac{1}{4w} \underline{b}^T \underline{\underline{A}}^{-1} \underline{b} \right) \bigg]^n  \bigg|_{z=0}  \, . \end{eqnarray}
Here, $\underline{b}$ stands for a column matrix constituted by the $b_l (\{ \theta \} )$ and $\underline{\underline{A}}$ stands for a square matrix with the elements $A_{l, l^\prime} (\{s\} , z)$.  We extract the limit $n\to 0$ by expanding (\ref{struct3}) in powers of $n$ and find, up to a minus sign,
\begin{eqnarray}
\label{struct4}
&& \frac{\partial}{2w \, \partial z}  \bigg[  \frac{1}{2w} \underline{b}^T \underline{\underline{A}}^{-1} \underline{b} + \ln \det (w \underline{\underline{A}})  \bigg]  \bigg|_{z=0}  
\nonumber \\
&& = \, \frac{1}{4w^2} \underline{b}^T \left( \underline{\underline{A}}^{-1} \right)^\prime \underline{b} + \frac{1}{2w} \frac{\left( \det ( \underline{\underline{A}}) \right)^\prime}{\det  ( \underline{\underline{A}})} \, .
\end{eqnarray}
In writing Eq.~(\ref{struct4}) we use the shorthand notations
\begin{eqnarray}
\label{struct5}
\left( \underline{\underline{A}}^{-1} \right)^\prime &=&  \frac{\partial}{\partial z} \underline{\underline{A}}^{-1} \Big|_{z=0} \, ,
\nonumber \\
 \left( \det \left( \underline{\underline{A}} \right) \right)^\prime &=&  \frac{\partial}{\partial z} \det \left(\underline{\underline{A}} \right) \Big|_{z=0} \, .
\end{eqnarray}
 The first term on the right hand side of Eq.~(\ref{struct4}) is a homogeneous function of the Schwinger parameters $s_p$ of degree $-1$, i.e., is behaves generically like $s^{-1}$. The second term is homogeneous in the $s_p$ of degree $0$, i.e., it goes generically like $s^0$. 

Now to the momentum-featuring part of our typical diagram depicted in Fig.~\ref{diagramStruct}. Suppose the diagram has $L$ closed loops. Then the momentum integrations result in a factor that is a homogeneous function of the Schwinger parameters of degree $-Ld/2$. In other words, the momentum integrations yield a generic factor $s^{-Ld/2}$. 

Next we turn to the integration over the Schwinger parameters. Suppose our diagram has $P$ propagators. Having each propagator represented in Schwinger parametrization, we have $P$ integrations over Schwinger parameters. These may be viewed as a generic factor $s^P$. Collecting, we find that the term proportional to $w^{-1}$ in Eq.~(\ref{struct4}) goes like $s^{P-Ld/2}$ and that the term 
proportional to $w^{-2}$ goes like $s^{P-Ld/2-1}$. By inspection one finds that the following topologic relations apply to our typical diagram: $L = P -V -1$ and $3V +4 = 2P$, where $V$ denotes the number of vertices. For the $w^{-1}$-term these relation lead to $P-L\frac{d}{2} = - \frac{d-6}{2} V +4 -d$, i.e., at $d=6$ this term goes generically like $s^{-2}$. Similarly, one finds a $s^{-3}$-behavior for the 
$w^{-2}$-term. Now we have sufficient information to single out the $\tau$-dependence of the terms. By a change of variables of the type $s\to s/\tau$ we learn that the $w^{-1}$-term is associated with a factor  $\tau^2$ whereas the $w^{-2}$-term features a $\tau^3$. 

For transition II we still have to deal with the integrations over the $\theta$'s. These, however, merely result in purely numeric factors. Harvesting the findings of the above reasoning, we deduce that the divergent parts susceptibilities have the general structure
\begin{eqnarray}
\label{trans1}
\chi^{(1)}_{\text{div}}&=& \frac{T \, \tau^2}{w} \, X(u)  \, ,  
 \\
\label{trans2}
\chi^{(2)}_{\text{div}} &=& \frac{T \, \tau^3}{w^2} \, Y(u) + \frac{T \, \tau^2}{w} \, X(u)   \, .
\end{eqnarray}
It is understood that Eq.~(\ref{trans1}) and Eq.~(\ref{trans2}) refer to transition I and transition II, respectively. The coefficients $X(u)$ and $Y(u)$ have a Laurent expansion of the form
\begin{eqnarray}
\label{struct7}
X(u) = \sum_{k=1}^\infty  \frac{X_k (u)}{\varepsilon^k}  &\text{with}&  X_k (u) = \sum_{m \geq k-1}  X_{k,m} u^m \, ,
\nonumber \\
\\
Y(u) = \sum_{k=1}^\infty  \frac{Y_k (u)}{\varepsilon^k} &\text{with}&  Y_k (u) = \sum_{m \geq k-1}  Y_{k,m} u^m \, ,
\nonumber \\
\end{eqnarray}
with $X_{k,m}$ and $ Y_{k,m}$ being numerical coefficients. As argued in Sec.~\ref{diagramatics:One-loopCalculation}, Eqs.~(\ref{trans1}) and (\ref{trans2}) have to be supplemented by terms varying as powers of $\Lambda$. This gives finally
\begin{eqnarray}
\label{trans1b}
&&\chi^{(1)}_{\text{div}}= \frac{T}{w} \Big[ A_0 (u) \Lambda^4 + A_1 (u) \Lambda^2 \, \tau  +  X(u) \, \tau^2 \Big]  \, ,  
 \\
\label{trans2b}
&&\chi^{(2)}_{\text{div}} = \frac{T}{w} \Big[ A_0 (u) \Lambda^4 + A_1 (u) \Lambda^2 \, \tau   + X(u) \, \tau^2 \Big]
\nonumber \\
&&  + \, \frac{T}{w^2} \Big[ B_0 (u) \Lambda^6 + B_1 (u) \Lambda^4 \, \tau + B_2 (u) \Lambda^2 \, \tau^2
+ Y(u) \, \tau^3 \Big]   \, ,
\nonumber \\
\end{eqnarray}
where $A_0 (u)$, $ A_1 (u)$ and so on are $u$ dependent coefficients.

\subsubsection{Renormalization and scaling}
In Sec.~\ref{opCorr} we argued that the order parameter correlation functions can be renormalized by the  renormalization scheme~(\ref{renScheme}). This scheme, however, is not sufficient to renormalize the susceptibilities. To remove the $\varepsilon$-poles encountered in Secs.~\ref{diagramatics:One-loopCalculation} and \ref{diagramatics:GeneralStructure}, one has to resort to an additive renormalization
\begin{eqnarray}
\label{addRen}
\mathcal{H}_{\text{ren}} \to \mathcal{H}_{\text{ren}} + \frac{1}{4} \, \chi^{(1),(2)}_{\text{div}} \int d^dx \, F_{i,j} (\brm{x}) F_{i,j} (\brm{x}) \, . 
\end{eqnarray}
Here $\mathcal{H}_{\text{ren}}$ stands for the renormalized Hamiltonian obtained form the original bare Hamiltonian $\mathcal{H}$ by applying the renormalization scheme~(\ref{renScheme}). We point out  that (\ref{addRen}) is not adequate to renormalize the free energy completely. As power counting shows, one also needs additive counter-terms of third and fourth order in $F_{i,j} $ as well as counter-terms containing derivatives of $F_{i,j} $. However, for our central task, i.e., for determining the scaling behavior of the susceptibilities, it is sufficient to consider (\ref{addRen}). Hence, we neglect the other just mentioned additive renormalizations. For our setup with the fixed external field $B$, (\ref{addRen}) implies for the free energy that 
\begin{eqnarray}
\label{addF}
\mathcal{F}_{\text{ren}} (B) \to \mathcal{F}_{\text{ren}} (B) + \frac{V}{2} \chi^{(1),(2)}_{\text{div}} B^2 \, .
\end{eqnarray}
According to definition~(\ref{replicaChi}) this implies
\begin{eqnarray}
\label{addChi}
&&\chi^{(1),(2)} ( \tau, w, g ) \to  \mathring{\chi}^{(1),(2)} \left( \mathring{\tau}, \mathring{w}, \mathring{g} \right) 
\nonumber \\
&&= \, \chi^{(1),(2)} (\tau, w, u, \mu ) + \chi^{(1),(2)} (\tau, w, u, \mu )_{\text{div}}
\end{eqnarray}
for the susceptibilities.

Now we are in good shape to analyze the scaling behavior of the susceptibilities. In order to reduce the use of indices and to keep the arguments as plain as possible, we carry out the following steps at the instance of $\chi^{(1)}$. At the end it will be straightforward to adapt our arguments to $\chi^{(2)}$. 

Just like the bare order parameter correlation functions, $\mathring{\chi}^{(1)}$ has to be independent of the inverse length scale $\mu$, i.e., it satisfies the identity
\begin{eqnarray}
\mu \frac{\partial}{\partial \mu} \mathring{\chi}^{(1)} \left( \mathring{\tau}, \mathring{w}, \mathring{g} \right) = 0 \, .
\end{eqnarray}
This identity is now taken as the origin of an RGE for $\chi^{(1)}$. Expressing the bare quantities through their renormalized counterparts one arrives initially at
\begin{eqnarray}
&&\left[ \mu \frac{\partial }{\partial \mu} + \beta \frac{\partial }{\partial u} + \tau \kappa \frac{\partial }{\partial \tau} + w \zeta \frac{\partial }{\partial w} \right] 
\nonumber \\
&& \times \, \left\{  \chi^{(1)} \left( \tau, w, u, \mu \right) +  \chi^{(1)} \left( \tau, w, u, \mu \right)_{\text{div}} \right\} =0 \, .
\nonumber \\
\end{eqnarray}
Here it is important to realize that all the individual terms appearing in the RGE have to be free of $\varepsilon$-poles. All terms associated with $\varepsilon$-poles must cancel order by order in the loop expansion. Taking into account the form of the Laurent expansion~(\ref{struct7}), the form of the Wilson function $\beta$ as given in (\ref{betau}) and that $\chi^{(1)}_{\text{div}} \sim \mu^{-\varepsilon}$ we obtain the RGE
\begin{eqnarray}
\label{rgeChi1}
&&\left[ \mu \frac{\partial }{\partial \mu} + \beta \frac{\partial }{\partial u} + \tau \kappa \frac{\partial }{\partial \tau} + w \zeta \frac{\partial }{\partial w} \right]  \chi^{(1)} \left( \tau, w, u, \mu \right)
\nonumber \\
&& = \frac{T \, \tau^2}{w} \, \overline{X}_1 (u) \mu^{-\varepsilon}
\end{eqnarray}
where $ \overline{X}_1 (u) \mu^{-\varepsilon} = (1 + u\partial_u) X_1 (u)$.  

Since the RGE~(\ref{rgeChi1}) is inhomogeneous, its solution is of the form
\begin{eqnarray}
\chi^{(1)} \left( \tau, w, u, \mu \right) = \chi^{(1)}_h \left( \tau, w, u, \mu \right) + \chi^{(1)}_p \left( \tau, w, u, \mu \right) \, ,
\nonumber \\
\end{eqnarray}
where $\chi^{(1)}_h$ is the general solution of the corresponding homogeneous equation and $\chi^{(1)}_p$ is a particular solution of the inhomogeneous equation. At the fixed point $u^\ast$ the method of characteristics gives for the homogeneous solution
\begin{eqnarray}
\label{solHom}
\chi^{(1)}_h \left( \tau, w, u, \mu \right) = \chi^{(1)}_h \left( \ell^{\kappa^\ast}\tau, \ell^{\zeta^\ast}w, u^\ast, \ell \mu \right) \, ,
\end{eqnarray}
where $\kappa^\ast = \kappa (u^\ast)$ and $\zeta^\ast = \zeta (u^\ast)$. This solution has to be complemented by a dimensional analysis, 
\begin{eqnarray}
\label{dimHom}
\chi^{(1)}_h \left( \tau, w, u, \mu \right) = \mu^{d-4} \chi^{(1)}_h \left( \mu^{-2}\tau,\mu^{-2}w, u, 1 \right) \, .
\end{eqnarray}
From Eqs.~(\ref{solHom}) and (\ref{dimHom}) we deduce that
\begin{eqnarray}
\label{totalHom}
\chi^{(1)}_h \left( \tau, w, u, \mu \right) = \ell^{d-4} \chi^{(1)}_h \left( \ell^{-1/\nu}\tau, \ell^{-\phi/\nu}w, u^\ast, \mu \right) \, .
\nonumber \\
\end{eqnarray}
Exploiting our freedom to choose the flow parameter, we set $\ell = |\tau |^\nu$. This choice yields
\begin{eqnarray}
\label{scalingHom}
\chi^{(1)}_h \left( \tau, w, u, \mu \right) =  |\tau |^{(d-4)\nu} f^{(1)} \left( w \, |\tau |^{-\phi} \right)
\end{eqnarray}
 with $f^{(1)}$ being a scaling function. Due to Eq.~(\ref{trans1}) we know that $f^{(1)} (x) \sim x^{-1}$. Thus, the homogeneous solution may be written as
\begin{eqnarray}
\label{scalingHomLead}
\chi^{(1)}_h \left( \tau, w, u, \mu \right) \sim \frac{T \, |\tau |^{t-2\nu}}{w} \, ,
\end{eqnarray}
where $t = (d-2)\nu + \phi$ is the conductivity exponent of the RRN. 

A particular solution is readily found by making an ansatz that is as similar to the inhomogeneity as possible. We obtain
\begin{eqnarray}
\label{partSol}
\chi^{(1)}_p \left( \tau, w, u, \mu \right)  = \frac{T \, \tau^2}{w} \, \frac{\overline{X}_1^\ast \, \mu^{-\varepsilon}}{2 - \varepsilon - 2/\nu + \phi /\nu} \, ,
\end{eqnarray}
where $\overline{X}_1^\ast = \overline{X}_1 (u^\ast)$.

Combining $\chi^{(1)}_h$, $\chi^{(1)}_p$ and the terms varying as a power of $\Lambda$ gives us the scaling behavior of $\chi^{(1)}$, viz.,
\begin{eqnarray}
\label{scalingTotalLead}
\chi^{(1)}  = \frac{T}{w}  \Big[ \Lambda^4 +  \Lambda^2 |\tau |^1 + |\tau |^2 \Big] + \frac{T\,  |\tau |^{t-2\nu}}{w} \, ,
\end{eqnarray}
where we have replaced nonuniversal coefficients, that might obscure the essential structure, by unity. Finally, we recall that our coarse grained parameter $w$ is proportional to the temperature. Ignoring the proportionality constant we thus get
\begin{eqnarray}
\label{scalingTotalLeadFinal}
\chi^{(1)}  =  \Lambda^4 +  \Lambda^2 |\tau | + |\tau |^2  +  |\tau |^{t-2\nu} \, .
\end{eqnarray}
The first terms correspond to the beginning of the small $\tau$ expansion of the regular part of $\chi^{(1)}$. The last term characterizes the leading behavior of the singular part of $\chi^{(1)}$. Note that, over all, the leading small $\tau $ behavior of $\chi^{(1)}$ is determined by its regular part. In passing we mention that the coefficients of the $|\tau |^2$ term and the singular term form a universal amplitude ratio that may be calculated with help of Eq.~(\ref{partSol})

An analysis similar to the one presented in the preceding paragraphs can be applied to $\chi^{(2)}$. Since only slight modifications are required, we simply state the result
\begin{eqnarray}
\label{scalingTotalChi2}
&& \chi^{(2)} = \Lambda^4 +  \Lambda^2 |\tau | + |\tau |^2 
\nonumber \\
&&
 + \,  \frac{1}{T}  \Big[ \Lambda^6 +  \Lambda^4 |\tau | +  \Lambda^2 |\tau |^2 +  |\tau |^3 \Big] 
\nonumber \\
&& + \, 
\frac{|\tau |^{2t-d\nu}}{T} \, \left[  1 + T \, |\tau |^{-\phi} \right]   \, ,
\end{eqnarray}
where we have once more replaced unimportant constants by unity. As above, the regular part gives the leading small $|\tau |$ behavior.

\section{Summary and concluding remarks}
\label{concludingRemarks}
In summary, we have studied the scaling behavior diamagnetic susceptibility $\chi^{(1)}$ and the mean square fluctuations of the total magnetic moment $\chi^{(2)}$ of large clusters as the Meissner (transition I) and the SG phase (transition II) are approached as $p \to p_c$ at low $T$. Our main results are summarized by the formulae
\begin{eqnarray}
\chi^{(1)}  &=&   \chi^{(1)}_{\text{reg}} +  |\tau |^{t-2\nu} \, ,
\\
\chi^{(2)} &=& 0
\end{eqnarray}
for transition I and
\begin{eqnarray}
\chi^{(1)} &=& 0 \, ,
\\
\chi^{(2)} &=&  \chi^{(2)}_{\text{reg}} + \frac{|\tau |^{2t-d\nu}}{T} \, \left[  1 + T \, |\tau |^{-\phi} \right]
\end{eqnarray}
for transition II. We remind the reader that $\tau$ measures the distance from the respective transition, $\tau \sim p_c -p$. $\nu$ is the percolation correlation length exponent and $t$ and $\phi$ are the conductivity and the resistance exponent, respectively, of the RRN. $\chi^{(1)}_{\text{reg}}$ and $ \chi^{(2)}_{\text{reg}}$ summarize the regular parts of the susceptibilities. These regular parts are very important and must not be neglected. In fact, they determine the leading small $\tau$ behavior of $\chi^{(1)}$ and $\chi^{(2)}$. As far as the leading small $T$ behavior is concerned, our results capture anticipated features of the susceptibilities. Typical for a diamagnetic susceptibility, $\chi^{(1)}$ approaches at transition I for $T\to 0$ a finite constant. $\chi^{(2)}$ on the other hand diverges at transition II as $T^{-1}$. That is because $\chi^{(2)}$ represents a paramagnetic rather than a diamagnetic susceptibility due to randomly frozen magnetic momenta in the spin glass phase.

We point out that our results hold to arbitrary order in $\varepsilon$-expansion. Using the powerful methods of renormalized field theory, we were able to explored general structural properties of the Feynman diagrams contributing to the susceptibilities. This allowed us to determine the scaling behavior of $\chi^{(1)}$ and $\chi^{(2)}$ to arbitrary order in perturbation theory.  

Our work reveals that the results by JLW are not entirely correct. On one hand, JLW overlooked the additive character of the renormalization of the susceptibilities that leads to regular contributions. On the other hand, JLW did not realize that the coupling constant $w_2$ associated with $T^2$ can enter the susceptibilities only in form of an irrelevant combination with $w$. This is true though and consequently exponents associated with $T^{-2}$ do not enter into the leading singular behavior of the susceptibilities.

Closing, we would like to mention interesting issues for future work. For example one could extend our work by investigating the diamagnetism beyond transition I and II, i.e., for the other transitions featured in the phase diagram of the RJN. One might look at the normal to superconducting transition as $p \to p_c$ away from the immediate vicinity of $T=0$ [cf.\ the critical surface (1) in Fig.~\ref{phaseDiag}]. Using our Gaussian propagator, this should be a feasible task. Another interesting subject is the role of vortex excitations, that we believe to be important in two dimensions. To address this question one has to develop an approach that avoids a linearization of the network equations. A potential strategy~\cite{foltin_stenull_janssen} is to devise a Villain type model~\cite{villain_75} for the RJN.

\begin{acknowledgments}
This work has been supported by the Deutsche Forschungsgemeinschaft via the Sonderforschungsbereich 237 ``Unordnung und gro{\ss}e Fluktuationen'' and the Emmy Noether-Programm. We are grateful to Georg Foltin for valuable discussions.
\end{acknowledgments}

\appendix
\section{The propagator}
\label{app:propagator}
In this Appendix we derive the Gaussian propagator as given in Eq.~(\ref{boldProp}). For notational simplicity we set $T=1$. 

The task is to solve the differential equation
\begin{eqnarray}
\label{prop1}
\left[  \tau \big( \vec{\lambda} \big) - \Big( \nabla - i \vec{\lambda} \cdot \vec{\brm{A}} \Big)^2  \right] G \big(  \brm{x} , \brm{x}^\prime , \vec{\lambda} \big) = \delta \big(  \brm{x} - \brm{x}^\prime \big) \, .
\end{eqnarray}
Here we use the shorthand notation $\tau (\vec{\lambda}) = \tau + w \vec{\lambda}^2$ introduced in Sec.~\ref{diagrammaticElements}.  Choosing for convenience the Landau gauge $\vec{\brm{A}} (\brm{x}) = \vec{B} (0, x, \brm{0}_\perp )$ , Eq.~(\ref{prop1}) takes on the form 
\begin{eqnarray}
\label{prop2}
&& \left[  \tau \big( \vec{\lambda} \big) -  \nabla^2 + 2i \omega  \big( \vec{\lambda} \big) x \partial_y +   \omega \big( \vec{\lambda} \big)^2 x^2 \right] G \big(  \brm{x} , \brm{x}^\prime , \vec{\lambda} \big) 
\nonumber \\
&&= \, \delta \big(  \brm{x} - \brm{x}^\prime \big) \, .
\end{eqnarray}
Inspired by Lawrie we rewrite the propagator as
\begin{eqnarray}
\label{prop3}
&&G \big(  \brm{x} , \brm{x}^\prime , \vec{\lambda} \big) = \exp \left[  i \frac{\omega  \big( \vec{\lambda} \big)}{2} (x+x^\prime) (y-y^\prime ) \right] 
\nonumber \\
&& \times \, \overline{G} \big(  \brm{x} , \brm{x}^\prime , \vec{\lambda} \big) \, ,
\end{eqnarray}
where $\omega  ( \vec{\lambda})$ is the cyclotron frequency given in Eq.~(\ref{frequency}). Inserting the propagator~(\ref{prop3}) into Eq.~(\ref{prop2}) we get
\begin{eqnarray}
\label{prop4}
&& \bigg[  \tau \big( \vec{\lambda} \big) -  \nabla^2 +i \omega  \big( \vec{\lambda} \big) \left[  \left( x - x^\prime \right) \partial_y -  \left( y - y^\prime \right) \partial_x  \right]
\nonumber \\
&& + \, \frac{ \omega  \big( \vec{\lambda} \big)^2}{4} \left[  \left( x - x^\prime \right)^2  +  \left( y - y^\prime \right)^2   \right] \bigg]  \overline{G} \big(  \brm{x} , \brm{x}^\prime , \vec{\lambda} \big) 
\nonumber \\
&&= \, \delta \big(  \brm{x} - \brm{x}^\prime \big) \, .
\end{eqnarray}
From Eq.~(\ref{prop4}) we deduce that $\overline{G} (  \brm{x} , \brm{x}^\prime , \vec{\lambda} )$ must be a function of the difference of the coordinates $\brm{x}$ and $\brm{x}^\prime$, i.e.,  $\overline{G} (  \brm{x} , \brm{x}^\prime , \vec{\lambda} ) = \overline{G} (  \brm{x} - \brm{x}^\prime , \vec{\lambda} )$. Hence, it is convenient to switch to momentum space via the Fourier transformation
\begin{eqnarray}
\label{prop5}
 \overline{G} \big(  \brm{x} , \vec{\lambda} \big) = \int_{\brm{k}} \widetilde{G} \big(  \brm{k} , \vec{\lambda} \big) \exp \left( i  \brm{k} \cdot \brm{x} \right) 
\end{eqnarray}
with $ \brm{k} = (p, q,  \brm{k}_\perp )$. Because  the system is rotationally invariant in the $x$-$y$-plane as well as in the hyperplane perpendicular to the $x$-$y$-plane, we anticipate the following form in momentum space: $ \widetilde{G} (  \brm{k} , \vec{\lambda} ) = \widetilde{G} ( p^2 + q^2,  \brm{k}_\perp^2 , \vec{\lambda} )$. Thus, Fourier transformation of Eq.~(\ref{prop4}) leads to
\begin{eqnarray}
\label{prop6}
 \bigg[  \tau \big( \vec{\lambda} \big) + \brm{k}^2  -  \frac{ \omega  \big( \vec{\lambda} \big)^2}{4} \left[  \partial_p^2 + \partial_q^2   \right]  \bigg] \widetilde{G} \big(  \brm{k} , \vec{\lambda} \big) = 1 \, .
\end{eqnarray}
For vanishing cyclotron frequency, this reduces to the well known equation for the propagator of the RRN. In Schwinger representation, the RRN propagator reads
\begin{eqnarray}
\label{prop7}
\widetilde{G} \big(  \brm{k} , \vec{\lambda} \big) = \int_0^\infty \exp \left[  -s \left(  \tau \big( \vec{\lambda} \big) + \brm{k}^2 \right) \right] \, .
\end{eqnarray}
However, our interest is not limited to the particular case $\omega  ( \vec{\lambda} ) = 0$. Hence, we generalize the solution~(\ref{prop7}) by making the ansatz,
\begin{eqnarray}
\label{prop8}
&& \widetilde{G} \big(  \brm{k} , \vec{\lambda} \big) =  \int_0^\infty ds \,  f(s)
\nonumber \\
&& \times \,  \exp \left[  -s   \left( \tau  ( \vec{\lambda} ) +  \brm{k}_\perp^2 \right) - g(s) \left(   p^2 + q^2 \right)  \right]  
\end{eqnarray}
with $f(s)$ and $g(s)$ being unknown functions of the Schwinger parameter $s$. We demand that
\begin{eqnarray}
\label{prop9}
&& \bigg[  \tau \big( \vec{\lambda} \big) + \brm{k}^2  -  \frac{ \omega  \big( \vec{\lambda} \big)^2}{4} \left[  \partial_p^2 + \partial_q^2   \right]  \bigg] \widetilde{G} \big(  \brm{k} , \vec{\lambda} \big) = - \int_0^\infty ds \, \frac{\partial}{\partial s}
\nonumber \\
&& \times  \, \Big\{ f(s)
  \exp \left[  -s   \left( \tau  \big(  \vec{\lambda} \big ) +  \brm{k}_\perp^2 \right) - g(s) \left(   p^2 + q^2 \right)  \right]  \Big\} \, .
\nonumber \\
\end{eqnarray}
For Eq.~(\ref{prop9}) to be satisfied, the unknown functions $f(s)$ and $g(s)$ have to satisfy the differential equations
\begin{eqnarray}
g^\prime (s) = 1 - \left[  \omega  \big(  \vec{\lambda} \big) g(s) \right]^2 \, ,
\\
- \frac{f^\prime (s)}{f(s)} = \omega  \big(  \vec{\lambda} \big)^2 g(s) \, ,
\end{eqnarray}
along with the boundary conditions $f(0) = 1$, $g(0) = 0$, and $g(s) \geq 0$. And, of course, the two functions must yield $f(s) \to 1$ and $g(s) \to 0$ for $\omega  (  \vec{\lambda} ) \to 0$. We find
\begin{eqnarray}
\label{prop10}
g (s) = \frac{\tanh \left(   \omega  \big(  \vec{\lambda} \big) s \right)}{ \omega  \big(  \vec{\lambda} \big)} \, ,
\\
f(s) = \frac{1}{\cosh \left(   \omega  \big(  \vec{\lambda} \big) s \right)} \, .
\end{eqnarray}
Inserting these results into our ansatz~(\ref{prop8}) we obtain $\widetilde{G} (  \brm{k} , \vec{\lambda} )$ as given in Eq.~(\ref{resGtilde}).


\begin{thebibliography}{}
\bibitem{footnote1}
To avoid spurious periodicities arising from the underlying lattice, the occupied bonds are assumed to be somewhat curved with a random length, c.f., Ref.~\cite{john_lubensky_85_86}.
\bibitem{footnote2}
In the RRN, the coupling constant $w$ decorating $\vec{\lambda}^2$ is proportional to the resistance $\rho$ of the occupied bonds.
\bibitem{footnote:reminder}
We would like to remind the reader that the constraint $\vec{\lambda} \neq \vec{0}$ has to be implemented properly in the continuum approximation.
\end{thebibliography}
\end{document}